%% file: ms.tex
\documentclass{sig-alternate-10pt}
\usepackage[utf8]{inputenc} % allow utf-8 input
\usepackage[T1]{fontenc}    % use 8-bit T1 fonts
\usepackage{hyperref}       % hyperlinks
\usepackage{url}            % simple URL typesetting
\usepackage{booktabs}       % professional-quality tables
\usepackage{amsfonts}       % blackboard math symbols
\usepackage{nicefrac}       % compact symbols for 1/2, etc.
\usepackage{microtype}      % microtypography
\usepackage{xcolor}         % colors
\usepackage{anyfontsize}
\usepackage{caption} 
\usepackage{siunitx}
\usepackage{graphicx,subcaption}
\usepackage{dblfloatfix}

\usepackage{enumitem}
\usepackage{multirow}
\usepackage{balance}

\newcommand{\xref}[1]{\S\ref{#1}}

\newcommand{\squishlist}{\begin{itemize}[itemsep=1pt,parsep=2pt,topsep=3pt,partopsep=0pt,leftmargin=0em, itemindent=1em,labelwidth=1em,labelsep=0.5em]}
\newcommand{\squishend}{\end{itemize}}
\newcommand{\squishenum}{\begin{enumerate}[itemsep=1pt,parsep=2pt,topsep=3pt,partopsep=0pt,leftmargin=0em,listparindent=1.5em,labelwidth=1em,labelsep=0.5em]}
\newcommand{\squishsubenum}{\begin{enumerate}[itemsep=1pt,parsep=2pt,topsep=0pt,partopsep=0pt,leftmargin=0em,listparindent=1.5em,labelwidth=1em,labelsep=0.5em]}
\newcommand{\squishenumend}{\end{enumerate}}

\newenvironment{todo-env}{\par\color{red}}{\par}
\newenvironment{help-env}{\par\color{blue}}{\par}
\newenvironment{ready-for-review}{\par\color{violet}}{\par}

\newif\ifcomments
\commentstrue % COMMENT THIS LINE TO DISABLE COMMENTS
\ifcomments
    \newcommand{\steve}[1]{\textcolor{cyan}{[SS: #1]}}
    \newcommand{\shyam}[1]{\textcolor{magenta}{[SG: #1]}}
    \newcommand{\ira}[1]{\textcolor{red}{[I: #1]}}
\else
    \providecommand{\steve}[1]{}
    \providecommand{\shyam}[1]{}
    \providecommand{\ira}[1]{}
\fi

\title{{Wireless earbuds for low-cost   hearing  screening}}

\author{Justin Chan,$^{\diamond 1}$ Antonio Glenn,$^{\diamond 1}$ Malek Itani,$^{1,2}$ \\Lisa R. Mancl$^3$, Emily Gallagher,$^4$ Randall Bly,$^{4,5}$ Shwetak Patel,$^{1,2}$ and Shyamnath Gollakota$^1$\\ \\
$^\diamond$Co-primary student authors\\
 $^1$Paul G. Allen School of Computer Science and Engineering, University of Washington\\
 $^2$Department of Electrical and Computer Engineering, University of Washington\\
 $^3$Department of Speech and Hearing Sciences, University of Washington\\
 $^4$Seattle Children’s Hospital and Research Institute\\
 $^5$Department of Otolaryngology — Head and Neck Surgery, University of Washington\\
}
\begin{document}

\maketitle

\input{abstract}

\input{intro-1}
\input{system}

\input{prior}

\input{two-stage}

\input{hardware}

\input{eval}
\input{related-1}
\input{limits}

\input{conclude}

\balance
\bibliographystyle{unsrt}
\bibliography{oae}

\end{document}

%% file: abstract.tex
\begin{abstract}
We present the first wireless earbud hardware that can perform hearing screening by detecting otoacoustic emissions. The conventional wisdom has been that detecting otoacoustic emissions, which are the faint sounds generated by the cochlea,  requires sensitive and expensive acoustic hardware. Thus,   medical devices for hearing screening   cost  thousands of dollars and are  inaccessible in low and middle income countries. We show that by  designing wireless earbuds using low-cost acoustic hardware and combining them with wireless sensing  algorithms, we can  reliably identify otoacoustic emissions and perform hearing screening. Our algorithms combine frequency modulated chirps with wideband pulses emitted from a low-cost speaker to reliably separate otoacoustic emissions from in-ear reflections and echoes. We conducted a clinical study with 50 ears  across two healthcare sites. Our study  shows  that the low-cost  earbuds detect hearing loss with 100\% sensitivity and 89.7\% specificity, which is comparable to the performance of a \$8000 medical device. By developing low-cost and open-source wearable technology, our work may help address global health inequities in hearing screening by democratizing these medical devices. 
\end{abstract}

%% file: intro-1.tex
\section{Introduction}
\vskip 0.15in

%It is estimated that 5.3% of the world’s population suffers from disabling hearing loss. Also, a disproportionate brunt of this problem falls on low- and middle-income countries (LMICs)1. Hearing loss can be especially harmful for neurodevelopment if untreated in early childhood. However, the impact of hearing loss may be mitigated when detected and treated early2. It is common practice for high-income countries to adopt guidelines for universal infant hearing screening using otoacoustic emission (OAE) or auditory brainstem response (ABR)3 testing. In spite of this, the test equipment remains expensive and costs thousands of dollars, which contributes to limited hearing screening in LMICs. In these countries, access to hearing assessment and equipment often requires travel to an urban setting and long wait times4,5.

\begin{figure}[t!]
\centering
    \includegraphics[width=.35\textwidth]{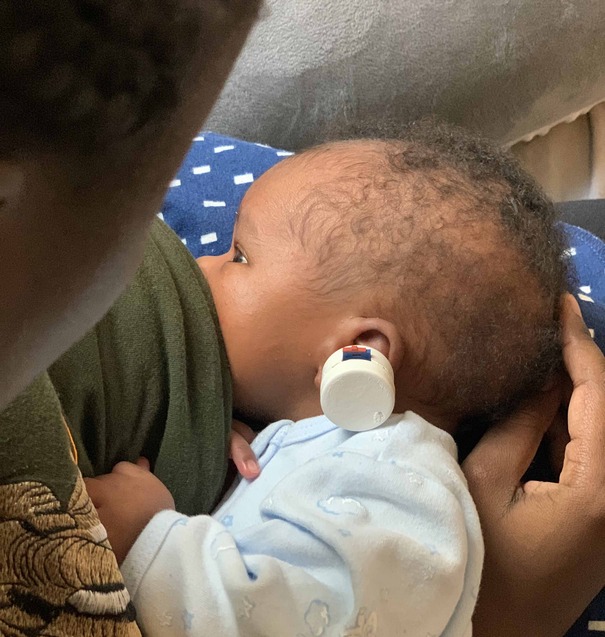}
    \vskip -0.05in
\caption{{\bf OAEbuds in use with an infant.} Our low-cost wireless earbud can perform  hearing screening by detecting otoacoustic emissions (OAE) from the cochlea.}
\vskip -0.15in
\label{fig:fig1}
\end{figure}

The World Health Organization estimates that 5.3\% of the world's population suffers from disabling hearing loss and 80\% of people who need hearing care live in  low and middle-income countries~\cite{who_hl,mcpherson2012newborn,swanepoel2010telehealth}. Hearing loss is particularly harmful for neuro-development if it is left undetected in early childhood~\cite{standard2,standard3,nbme22}. As a result, high-income countries  have guidelines for universal infant hearing screening --- the Joint Committee on Infant Hearing, the  American Academy of Pediatrics and  the Centers for Disease Control and Prevention all recommend  universal hearing screening~\cite{standard2,standard3,standard1,standard4} that is now implemented across almost all states, communities and hospitals in the United States~\cite{patel2011universal,cdc}.

Since the neonatal population cannot provide behavioural response to conventional audiometry tests~\cite{app3,app1,app2,walker2013audiometry}, existing newborn hearing screening technologies instead use the sounds generated by a healthy cochlea called otoacoustic emissions 
(OAE)~\cite{abdala2001distortion,chan2022inner}. While we think of the ear as a biological organ that receives sounds like a microphone, a healthy cochlea, the part of the inner ear responsible for converting sound waves into electronic impulses for the brain, also generates sounds.   These emissions  are created when the cochlea's sensory hair cells vibrate in respond to external sounds~\cite{abdala2001distortion,thompson2001universal,torre2003distortion}. So, we could pick up these faint sounds and use their absence to detect hearing loss.

The challenge is that detecting  these faint  sounds  emitted from the cochlea requires sensitive acoustic hardware and medical devices  that are expensive (5000-8000 dollars)~\cite{product1,product2}. As a result,  there is limited to no-hearing screening in low and middle-income countries~\cite{india1,kenya1}.  Further, in rural and resource-limited settings, getting access to hearing assessment  may often require travel to an urban setting and  long wait times, significantly limiting the accessibility of hearing care~\cite{mcpherson2012newborn,swanepoel2010telehealth,nbme22}.

\begin{figure}[t!]
\centering
    \includegraphics[width=.2\textwidth]{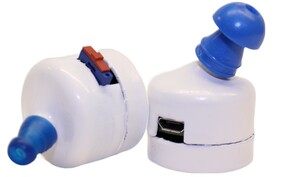}
    \includegraphics[width=.2\textwidth]{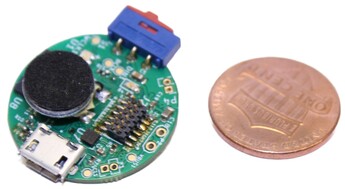}\\
    (a)\ \ \ \ \ \ \ \ \ \ \ \ \ \ \ \ \ \ \ \ \ \ \ \ \ \  (b)
    \vskip -0.05in
\caption{{\bf OAEbuds hardware.} (a) The 3D-printed enclosure with pediatric and adult earbud tips. (b) OAEbud circuit board  beside a penny for size comparison.}
\vskip -0.15in
\label{fig:oaebuds}
\end{figure}

We present OAEbuds, the first wireless earbud design  for low-cost hearing screening. Our hardware-software system reliably detects otoacoustic emissions using low-cost acoustic hardware, while being in the form-factor of a wireless earbud. The earbud hardware is designed to work across a wide demographic from new-borns to adults.\footnote{OAE testing is not limited to just newborns but is also used as part of clinical care in older kids and adults~\cite{asha.org}.} Our design  streams the digital acoustic data via Bluetooth to a nearby smartphone which is then used for processing the signals and displaying the test results. 

%Finally, we design algorithms to detect OAEs in the presence of some background noise.

\begin{figure}[t!]
%\vskip -0.15in
\centering
    \includegraphics[width=.47\textwidth]{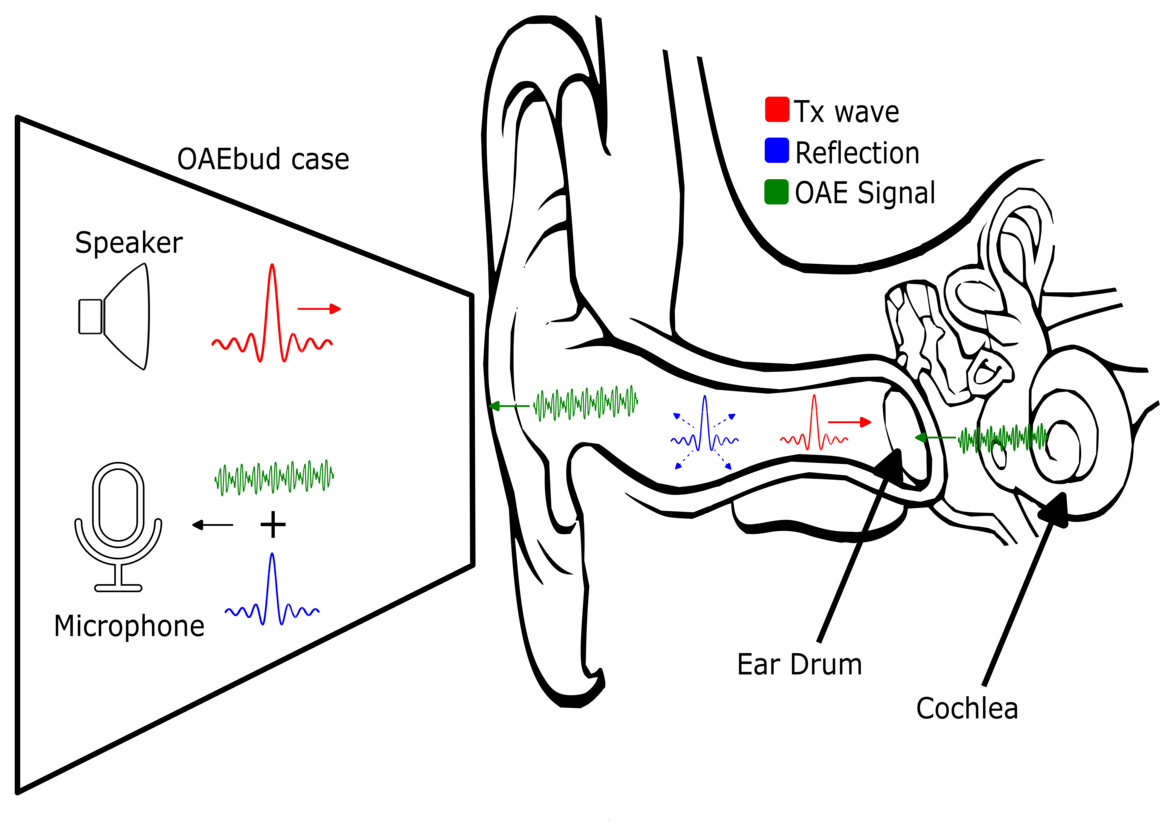}
  \vskip -0.15in
\caption{{\bf In-ear signal propagation.} During the measurement, the OAEbud plays a broadband transmit (Tx) pulse  to stimulate the cochlea to emit an OAE signal. The signal received by microphone is a superposition of 1) unwanted reflections from the ear canal, eardrum, and within the case and 2) the OAE signal.}        
\vskip -0.15in
\label{fig:cochlea}
\end{figure}

There are two key technical challenges in achieving this design with low-cost acoustic components. First, since speaker hardware components are bulky, it is challenging to incorporate the two-speaker design that has been proposed in recent work~\cite{nbme22} into the form-factor of a wireless earbud. Our experiments in~\xref{sec:existing} show  that transmitting the dual-tone signals used in~\cite{nbme22} on a  single low-cost speaker hardware introduces    nonlinearities that in turn creates inter-modulation tones  that can be confused for OAEs. Second,  when an acoustic signal is sent into an ear canal, it first gets reflected and creates echoes  not only inside  the earbud case but also the ear drum and  the walls of the ear canal, before arriving at the cochlea (see Fig.~\ref{fig:cochlea}). To accurately identify OAEs, it is important to determine when the reflections and echoes of the input stimuli end and the OAEs begin. 

We design a  two-step protocol that uses wireless sensing techniques to address the above challenges using a single low-cost speaker in our wireless earbuds.

\squishlist
\item {\bf Reflection time estimation.} In the first step, we send  frequency modulated continuous wave (FMCW) signals as input stimuli. These signals get reflected back from the earbud case, the ear drum and the ear canal which are captured at the  microphone. Since OAE signals are faint and non-linear, they do not create  linear FMCW reflections. So we can  perform FMCW processing to estimate the time-delays corresponding to the reflections and echoes and determine the duration after which their power  reduces below a preset threshold (see~\xref{sec:reflections}). 

\item {\bf OAE signal extraction.} In the second step,  we transmit a train of wideband pulses  from the earbud speaker. Since the travelling sound wave traverses more slowly in the cochlea~\cite{speed}, the otoacoustic emissions  still arrive delayed in time after the  reflections and echoes of the input stimuli. To extract these  signals, we first reduce  reflections by only considering the signals that arrive after the duration estimated in the previous step. We then synchronize the responses across multiple wideband pulses, combine them to improve the SNR of OAE signals and detect them using our  earbud system (\xref{sec:oae}).
\squishend

We designed an open-source  wireless earbud hardware shown in Fig.~\ref{fig:oaebuds} that is capable of transmitting the above signals from its speaker and wirelessly streaming  the microphone audio.  We designed our OAEbuds hardware using open source eCAD software, outsourced fabrication and assembly (\$28.3 per unit), and 3D printed the enclosures in-house. The earbud is designed to support multiple ear tip sizes that allows it to snugly fit for both new-born infants and adults. The battery in the earbud can be recharged via a USB connection within 3~hours. Our evaluation shows that on a single charge, the earbud can be used to perform up to 91 hearing tests.

We perform a clinical study  on 50  ears from 26 pediatric and adult patients across two different healthcare sites. We perform testing with both our earbud device as well as an FDA-cleared medical device that performs OAE detection. For the tested patients, the attending clinicians determined the ground truth for hearing loss using the  patient’s  hearing screen, audiograms, diagnostic auditory brain response, and  clinical history. Our evaluation shows that OAEbuds achieves a sensitivity of 100\%  and specificity of
89.7\%  in screening for hearing loss. In comparison, the FDA-cleared medical device achieves a sensitivity of 83.3\% 
 and specificity of  92.1\%.  Our   techniques  also  improve the area under the curve (AUC) from 0.847 to 0.950 over existing OAE  algorithms. Finally,
 our system can output a `pass' or  `refer' result for hearing screening in under 70~seconds.

\vskip 0.05in\noindent{\bf Contributions.} We make the following contributions. 
\squishlist
\item We design the first wireless earbuds to achieve low-cost hearing screening by detecting otoacoustic emissions. 

\item  We introduce a two-step protocol  that combines FMCW signals with wideband pulses to separate reflections and echoes from OAEs while using a single low-cost speaker. 
%\item We use multi-microphone  processing  algorithms to address ambient noise, showing that contrary to popular belief, OAEs can be extracted outside of quiet  settings.

\item We perform a clinical study that shows  our low-cost wireless earbud  detecting hearing loss with accuracies similar to a \$8000
FDA-cleared medical device.
\item Finally, we will make our code and hardware  open source to
help with adoption across the target settings. 
\squishend
\

\vskip 0.05in\noindent{\bf Comparison to prior work.} 
The closest to our work is recent work~\cite{nbme22} that uses the two speakers in a wired earphone. It transmits a different  frequency tone from each speaker  and uses an additional microphone that is placed next to the ear to create a smartphone attachment.  This  work  has multiple   constraints that limit its adoption in the target use-cases.  1) It uses a wired earphone and external microphone that are connected to a smartphone. Since it uses the smartphone's ADC, DAC and AGC, it requires manual calibration for each  smartphone model, which is challenging to generalize.  2) It uses two frequency tones that are transmitted from  two different speakers and looks for inter-modulation between the tones to detect OAE. The challenge is that it is difficult to  incorporate two speakers pointing into the ear-canal in a wireless earbud form factor and hence the techniques used in this prior work cannot be used for wireless earbuds (see~\xref{sec:existing}). 3) The various hardware components  are attached using plastic tubing  and glue which make it unreliable and difficult to scale and introduces a DIY aspect to the system. In an informal survey of clinicians in an African  (anonymized) country  conducted by the authors, participants noted that this DIY-aspect  could translate to lowered patient confidence in  the care received at the clinic. A low-cost yet high-tech device would be required to achieve wider adoption by clinicians. Our paper addresses the above limitations   and designs the first wireless earbuds for low-cost hearing  screening.  Compared to DIY devices, since our wireless earbud is more integrated while being low-cost, it may help broaden the adoption of our   hearing screening tool.

%% file: system.tex
\section{System design}
\vskip 0.15in
We first describe existing approaches to OAE sensing and their limits. We then present  our two-step protocol  to estimate the reflection time and extract  OAE signals. Finally, we present our  low-cost earbud hardware.

%% file: prior.tex
\subsection{Existing OAE approaches}\label{sec:existing}
\vskip 0.15in

The challenge with reliably detecting OAEs is identifying them in the presence of much stronger in-ear reflections. There are two key prior approaches.

\begin{figure}[t!]
  \centering
    \includegraphics[width=.4\textwidth]{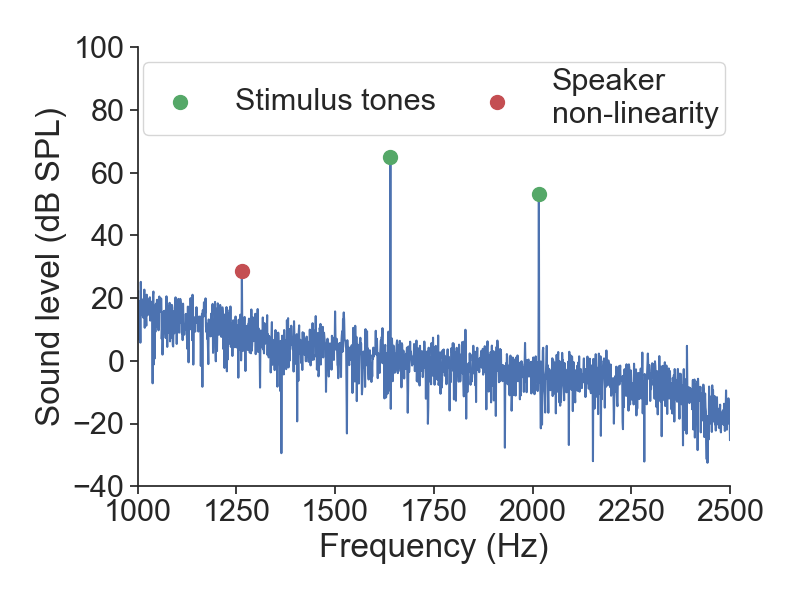}
    \vskip -0.15in
    \caption{{\bf Challenge of existing OAE approaches on a single-speaker earbud design.} Sending two stimulus tones $f_1$ and $f_2$ through a single speaker setup to elicit OAEs creates a hardware non-linearity at $2f_1-f_2$, which can be  stronger than the OAE signal at that frequency.}
    \label{fig:dpoae}
    \vskip -0.15in
\end{figure}

\squishlist
\item {\bf DPOAEs.}  Distortion-product otoacoustic emissions (DPOAEs) address the reflection problem by using intermodulation. In particular, the cochlea is stimulated by sending two tones $f_1$ and $f_2$. Given the nonlinear response of the basilar membrane within the cochlea, it generates a nonlinear intermodulation tone at the frequency $2f_1-f_2$~\cite{nbme22}.  Since reflections and echoes do not cause new frequency tones, this dual-tone approach can be used to separate OAEs from in-ear reflections. 
The challenge with deploying the DPOAE protocol on a single-speaker system is that hardware components introduce  nonlinear intermodulation distortion at frequencies which are linear combinations of  $f_1$ and $f_2$,   $k_1f_1 + k_2f_2$, where $k_1$ and $k_2$ are arbitrary integers. When $k_1=2$ and $k_2=-1$, this matches the  DPOAE signals  produced by the cochlea. These non-linearities are more prominent for low-cost speaker hardware. Fig.~\ref{fig:dpoae} shows the amplitude of the intermodulation produced by a single low-cost speaker (Knowles SR-32453-000, \$4.42)  when sending two tones 1640 and 2016~Hz at  65/55~dB SPL. The figure shows that the unwanted intermodulation component has a sound level of 28~dB SPL; in comparison  the typical range of DPOAEs  is 5--25~dB SPL~\cite{nbme22}. As a result, prior work~\cite{nbme22} uses a two-speaker system to separately send the $f_1$ and $f_2$ tones. Since wireless earbuds are generally constrained to only  a single speaker per bud due to size constraints, it is challenging to use the DPOAE protocol on such a low-cost  hardware.

\item {\bf TEOAEs.} The transient-evoked otoacoustic emission (TEOAE) protocol extracts the OAEs in the presence of in-ear reflections and echoes using a single speaker. Here a   short biphasic click sequence is repeatedly sent, with a polarity and amplitude pattern of $\{1,1,1,-3\}$ \cite{kemp_acoustic_1986} . The key insight behind this protocol is that the amplitude of the reflections are linearly related to the amplitude of the transmitted clicks. So the responses of all four clicks can be summed to cancel the reflections caused by the eardrum. However since OAEs are non-linear in nature, they would not be canceled by this addition operation. 
The challenge with deploying this protocol on a low-cost system, is that 1) the clicks need to be perfectly synchronized and phase-aligned, so that the reflections are cancelled out. Without exact alignment, there will continue to be residual energy caused by imperfect cancellation which will make it difficult to measure the OAEs. 2) Low-cost speaker hardware also introduces non-linearities in polarity and amplitude resulting in imperfect cancellation. Our evaluation in~\xref{sec:benchmark} shows that this leads to degraded performance. 
\squishend

 While other methods for eliciting OAEs have been proposed in the literature~\cite{neumann1994chirp,chan2000test}  they are not used in practice given uncertainty about their reliability. 

% using signals such as chirps or tones, they are not, to the best of our knowledge, used in practice in the clinic or implemented on commercial OAE devices. As such, using these methods would likely be met with more resistance from the clinical community, and a lower likelihood of adoption. Further, it would not be straightforward to compare the OAEs elicited from these types of signals to the calibrated DPOAE and TEOAE signals measured on conventional medical devices.

%% file: two-stage.tex
\subsection{Our two-stage protocol}
\vskip 0.15in

Instead of relying on the linearity of the acoustic hardware, we separate the in-ear reflections from OAEs in the time domain. At a high level, we first estimate the time delay at which reflections from both the ear and the enclosure drop below a particular threshold. We then detect the OAEs over the remaining time duration.

\begin{figure}[t!]
%\vskip -0.15in
\centering
    \includegraphics[width=.47\textwidth]{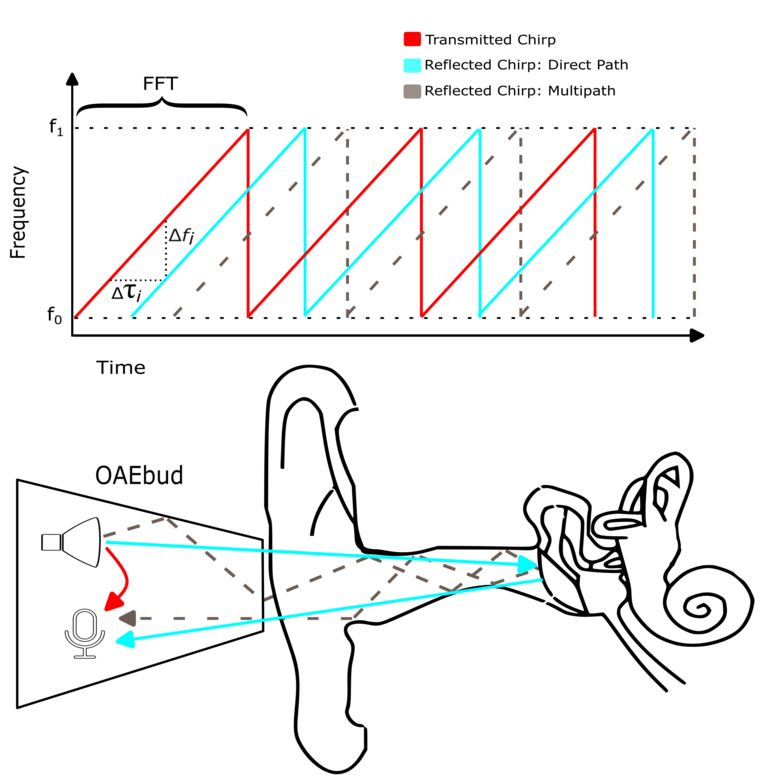}
  \vskip -0.15in
\caption{{\bf FMCW processing to calculate the time of arrival for reflections from the ear canal.} The OAEbud transmits a chirp into the ear canal, and record the reflections from the ear and enclosure. It then performs an FFT over the chirp duration to estimate the frequency shift $\Delta f_i$ and time delay $\tau_i$ for the $i^{th}$ reflection from the ear. This estimate is averaged across three chirps.}
\vskip -0.15in
\label{fig:chirpFMCW}
\end{figure}

\subsubsection{Reflection time estimation}\label{sec:reflections}
\vskip 0.15in

The time delay when reflections diminish will differ from one ear to another due to differences in anatomical structures such as ear canal diameter which increases with age. Further, hair and debris  can change the reflection profile significantly. At a high level,  we send an FMCW signal to estimate when reflections from the case and other parts of the ear diminish beyond a certain threshold. Our algorithm then uses the remainder of the recording after this time estimate to measure OAEs.

Although FMCW signals could be used to estimate the length of an individual ear canal, and convert that to a time delay at which reflections diminish, we find that in practice there is a significant amount of echos caused by reflections from the case, ear drum and ear canal that result in a large delay spread, much larger than the time of flight measurement for a typical ear canal length of 2.5~cm. An analogy to this would be that if one shouts in an empty cave, it can take several seconds for all the echos to diminish due to the significant reflections that occur from the cave walls. Further we note that the speed of sound is slower in the cochlea, meaning the OAE would take a longer time to return compared to if the pulse were only sent into the air medium, and this contributes further to a large delay spread~\cite{speed}.

We send a chirp with linearly increasing frequency from $f_0$ to $f_1$ where the frequency at a given time $t$ is denoted as $f(t) = f_0 + \frac{Bt}{T}$, where $B$ and $T$ are the bandwidth and duration of the chirp. The phase is computed by integrating $f(t)$ over time, resulting in the function: $\phi(t) = 2\pi(f_0t+B\frac{t^2}{2T})$. The signal that is then transmitted in the time domain is defined as $x(t) = cos(\phi(t))$, as shown in Fig.~\ref{fig:chirpFMCW}. 

Each of the echoes from the ear canal are delayed chirps that arrive at the microphone as the received signal, $y(t)$, that is a combination of all echoes. To estimate the multipath profile of the ear canal, we multiply the receiver signal with the  transmitted signal, $x(t) y(t)$. 
Using the trigonometric identify, $2cos(A) cos(B) = cos(A-B) + cos(A+B)$ and filtering out the high frequency term, $cos(A+B)$, we can translate the time delays of each of the echoes into  frequency shifts between the transmitted and  received chirps.  Note that in contrast to radio signals that have both I and Q components, acoustic signals operate in the real space. So instead of using a downchirp, we multiply the received cosine signal with the transmitted signal and apply a low-pass filter.

In Fig.~\ref{fig:chirpFMCW}, we show a plot of the transmitted FMCW signal, along with several reflections in the frequency domain, each with its own time delay $\tau_i$ for the $i^{th}$ reflection. In order to determine individual time delays $\tau_i$ when the reflections end, we compare the differences in frequencies between the transmitted and reflected signals. Specifically, a time delay $\tau_i$  will result in a frequency shift of  $\Delta f_i$ for the reflected signal from the transmitted signal and can be computed as follows:
$$\tau_i = \frac{\Delta f_iT}{f_1-f_0} $$

To obtain a precise  resolution for our reflection time estimate, we send an FMCW signal that is close to the maximum bandwidth allowable by the sampling rate of our system. As the maximum sampling rate of our OAEbuds hardware is 31250~Hz, we send an FMCW signal with a bandwidth of 5 to 15~kHz so that the upper frequency is close to the Nyquist frequency of 15625~Hz. The time resolution of an FMCW system is,  
$\frac{1}{2B}$
, where $B$ is the bandwidth. This corresponds to a time resolution of 0.05~ms when $B$=10~kHz. We set the length of our signal based on the maximum number of samples that can be stored in our hardware's memory. In our system we use a 200~ms FMCW signal which corresponds to 6250 samples.

% TODO
% $$$$ % TODO
% TODO

\begin{figure}[t!]
  \centering
    \includegraphics[width=.45\textwidth]{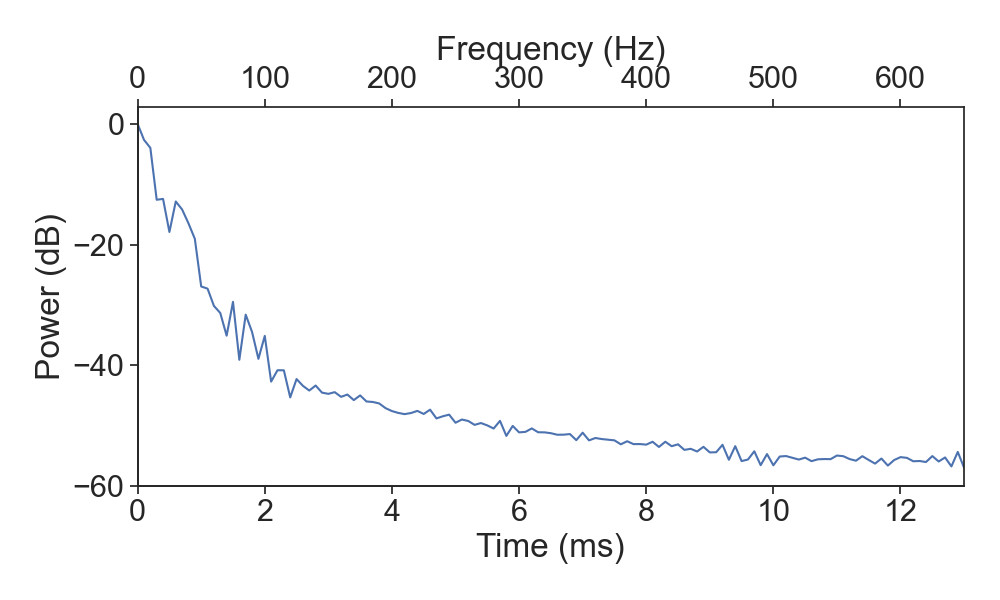}
    \vskip -0.15in
    \caption{{\bf Estimating when reflections diminish.} By measuring the frequency shifts from reflections of a FMCW signal transmitted into the ear canal, we can estimate the time delay $t_D$ after which reflections diminish to a predetermined power threshold.}
    \label{fig:delay}
    \vskip -0.15in
\end{figure}

Fig.~\ref{fig:delay} shows the result of this processing  in a normal adult ear. We can observe a peak in the zeroth bin that corresponds to the incident chirp, and peaks at subsequent bins that correspond to reflections arriving at increasing time delays. To minimize the interfering effects of reflections in our OAE measurement, we select a time delay that corresponds to the frequency shift where the power level diminishes below a preset power threshold. In our implementation, if the power level of a frequency bin decreases below 55~dB from the power of the incident signal, we use the time delay, $t_D$,  corresponding to that frequency bin. If such a  bin cannot be found, a default delay value of 12~ms is used.

We note two key points about our earbud system.
\squishlist
\item Performing this estimation  is important particularly given the tonotopic geometry of the cochlea (Fig.~\ref{fig:cochlea}) where the high frequency OAEs exit the cochlea first, followed by the low frequencies. The cochlea has a coiled shape where the beginning of the coil responds to high frequency sounds, while the inner most curled part of the coil responds to low frequency sounds. As such, when a stimulus pulse is sent into the cochlea, it is the high frequency OAEs that exit first, and it is these frequencies that would  also be most affected by the reflections. By setting a time delay that is too low, there is a risk that the reflections will be confused for the OAEs, whereas setting the delay too high may only result in a measurement over the low frequency OAEs, but few of the high frequency OAEs. Our algorithm allows us to minimize the power of reflections while increasing the power of the OAE signals.
\item Interestingly, the ear canal creates a closed enclosure that can create a large number of strong echoes. As a result, while the length of the ear canal is only around 2.5 cm~\cite{zemplenyi1985optical}, which translates to an acoustic round trip time of  0.15~ms, given the large number of echoes created within the ear canal, we can have reflections as shown in Fig.~\ref{fig:delay} that arrive even at 5-10~ms. This emphasizes the need for using a system that computes the time-delays for the in-ear reflections which can be much longer than the time it takes to traverse the ear canal. We also note that we compute the above time delay by averaging the values across three continuous  FMCW chirps.
\squishend

\subsubsection{OAE signal extraction}\label{sec:oae}
\vskip 0.15in
After the time delay has been identified, our system  needs to  reliably measure the faint  otoacoustic emissions that are as low as -10 to 30~dB SPL using low-cost microphones that would not have the same sensitivity of the high-end expensive microphones in medical devices. To extract these faint OAE signals, at a high level, we combine the OAE responses across multiple pulses to increase the SNR of OAEs. This can be challenging especially given that the target population of this test is young infants who may move, fidget or otherwise cause noise throughout the measurement. Our measurement should also be able to reliably distinguish between periods of noise caused by the patient and legitimate OAE signals, as an incorrect classification would result in an inaccurate measurement or an overly lengthy measurement that would result in patient discomfort. 

In the rest of this section, we first describe our transmission scheme and then describe the various steps needed to extract the OAE signals.

\begin{figure}[t!]
%\vskip -0.15in
\centering
    \includegraphics[width=.49\textwidth]{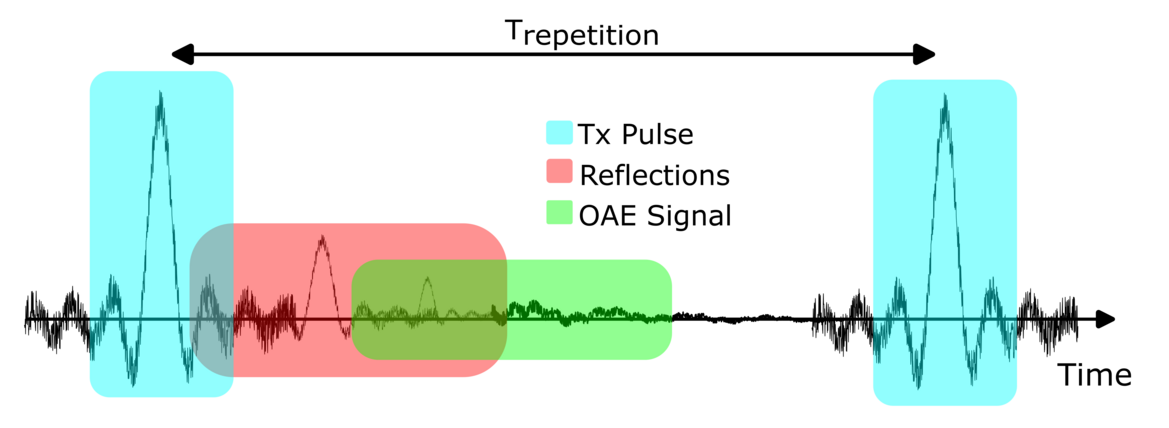}
  \vskip -0.15in
\caption{{\bf OAEbuds pulse transmission scheme.} A pulse of $500~\mu$s with a bandwidth from 0--5~kHz is transmitted every 20~ms into the ear to cover the range of frequencies important for hearing screening. The recorded signal consists of reflections of the input stimuli from the ear canal which  overlap with the OAE signal.}
\vskip -0.15in
\label{fig:pulse_sequence}
\end{figure}

\vskip 0.05in\noindent{\bf Pulse transmission scheme.} We transmit a sequence of short  $500~\mu$s pulses and apply a brick-wall filter with a bandwidth from $f_{start}=0~kHz$ to $f_{end}=5~kHz$, to cover the full range of frequencies of clinical interest in hearing screening (Fig.~\ref{fig:pulse_sequence}). We use a sampling rate of 15625~Hz, which corresponds to 8 samples to represent the pulse. We multiply the pulse with a hamming window to reduce the effect of ringing. Each of the pulses are separated by a gap of 20~ms, to ensure that the OAEs have enough time to arrive at the microphone. We  perform these measurements over the course of 67.5 seconds, which corresponds to approximately 3300 pulses.

\vskip 0.05in\noindent{\bf Decoding algorithms.} We describe the various steps to combine the OAE responses across pulses and extract higher SNR OAE signals.

{\it Step 1. Pulse synchronization.} The first step of our algorithm is to establish synchronization with the start of the pulse. To do this, we perform cross correlation of the first one second window with the transmitted pulse, and look for peaks with a minimum peak prominence set to 0.3e8. If we do not find such a peak in this window, we proceed to the next one second window. We note that we may not find peaks within the first one second window if the user initially begins the measurement outside the ear where the amplitude of the reflections will be lower compared to in the ear. Once the start of a pulse can be found, we can add a fixed offset of 20.5~ms to find the start of the next pulse. We also use this windowed approach instead of performing cross correlation over the entire one minute recording as it allows for real-time computation of the OAEs over the course of the measurement. In other words, our algorithm is able to incrementally compute the OAE result over windows of one second intervals, and provide continuous feedback to the user about the OAE results, and whether the environment is too noisy. This will allow the user to be able to recognize possible problems in probe fit or environmental noise in real-time instead of having to wait until  the entire one minute measurement is complete.

{\it Step 2: Noise detection.} After we have a set of peaks corresponding to the start of all the pulses in a 1-second window, our next step is to determine which pulses are  usable for subsequent processing and identify pulses that have been affected by noise in the environment. To do this, we apply a sliding correlation window over batches of four pulses and calculate the correlation between each of the adjacent pulses. We sum the calculated correlation values and divide it by the sum of the received power within that batch. This produces a normalized value that is invariant to pulse amplitude differences across batches. If this normalized value is above 0.95, we consider that batch to be usable for subsequent measurement, else we discard that batch. This allows us to discard specific OAE signals within the overall measurement that have been corrupted by noise.

{\it Step 3: Combining OAE responses across pulses.} For all usable batches in a given window, we look for OAE responses using the time delay  $t_D$ computed in the previous section. We  use  the window size of $t_P-t_D-t_{guard}$ where $t_P$ is the gap between consecutive pulses and $t_{guard}$ is a guard period which we set to 1~ms. In other words, we look for OAEs starting from the  time when the reflections have diminished up till the start of the next pulse, minus a small guard period. We then average the power of all odd numbered pulses to compute, $P^{OAE}_{odd}$ and all even numbered pulses for  $P^{OAE}_{even}$. We then compute the signal and noise power, $P_{signal}$ and  $P_{noise}$ in the time domain as follows:
$$P_{signal} = \frac{P^{OAE}_{odd}+P^{OAE}_{even}}{2}, P_{noise} = \frac{\lvert P^{OAE}_{odd}-P^{OAE}_{even}\rvert}{2}$$
To obtain the SNRs across 1 to 5~kHz, we convert the above signals to the frequency domain by performing an FFT. We repeat the above process for each frequency band by taking an average across adjacent  frequency bins. Specifically we perform an average over the following bands: 750--1250~Hz, 1250--1750~Hz, 1750--2500~Hz, 2500--3500~Hz, and 3500--4500~Hz. The result of this step is a set of SNR measurements for each frequency band. 

\vskip 0.05in\noindent{\bf Hearing testing algorithms.} Finally, we describe the algorithms we run while  performing the hearing test.

\vskip 0.05in{\it  a) Computing the pass/refer result.} To calculate the `pass' or `refer' screening result, as with existing medical devices, we determine if at least 2/3 of the  5 frequency bands are above a preset threshold (8~dB in our case). These parameters were determined by performing  a parameter sweep over different number of frequency bands and SNR thresholds to determine values that optimize our clinical performance. Further, we check that the absolute sound level of the signal component is above a preset value of -10~dB SPL, which is typically regarded as the minimum sound level of an OAE. This ensures that spurious reflections or noise that were not discarded during previous filtering steps are not mistaken as OAEs. Additionally, if the average noise level across the frequency bands exceeds 6~dB SPL, we mark the measurement as noisy.

\begin{figure}[t!]
%\vskip -0.15in
\centering
    \includegraphics[width=.4\textwidth]{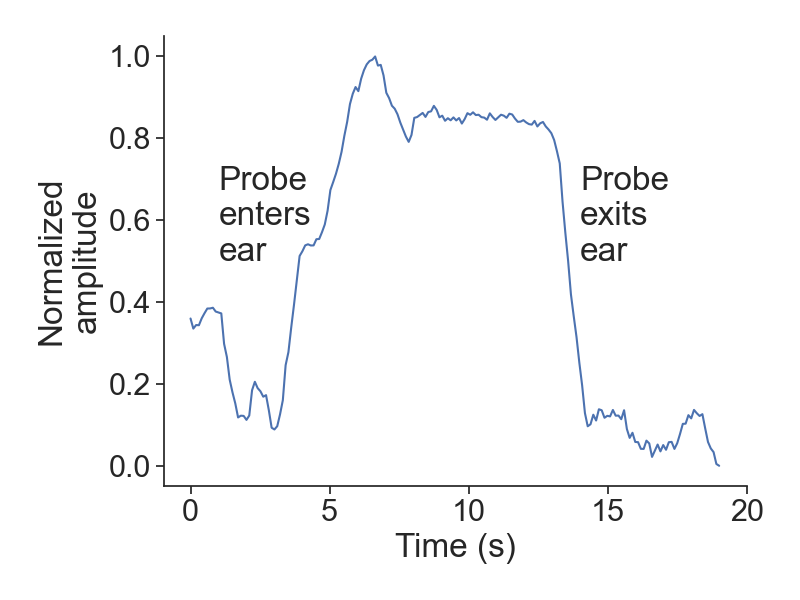}
  \vskip -0.1in
\caption{{\bf Checking if the probe is in the ear.} By measuring the amplitude of a chirp at the 200~Hz frequency during the beginning of a measurement, we can detect whether the ear probe is inside or outside the ear.}        
\vskip -0.15in
\label{fig:tone}
\end{figure}

\vskip 0.05in{\it b) Determining if probe is in the ear.} To determine when the measurement can begin, our system performs a check for whether the probe is probably placed in the ear. To do this, we send a sequence of 20~ms chirps from 100 to 5500~Hz and measure the amplitude of the frequency response to determine if the probe has formed a snug fit with the ear. We find that the frequency at 200~Hz is representative of whether the probe is outside or inside the ear (Fig.~\ref{fig:tone}). If the average sound level in this frequency range exceeds a predefined threshold for 50 chirps (1~s), we mark the probe tip as being in the ear and begin the  measurement.

%% file: hardware.tex
\subsection{Hardware design}
\vskip 0.15in

\begin{figure}[t!]
\centering
    \includegraphics[width=.49\textwidth]{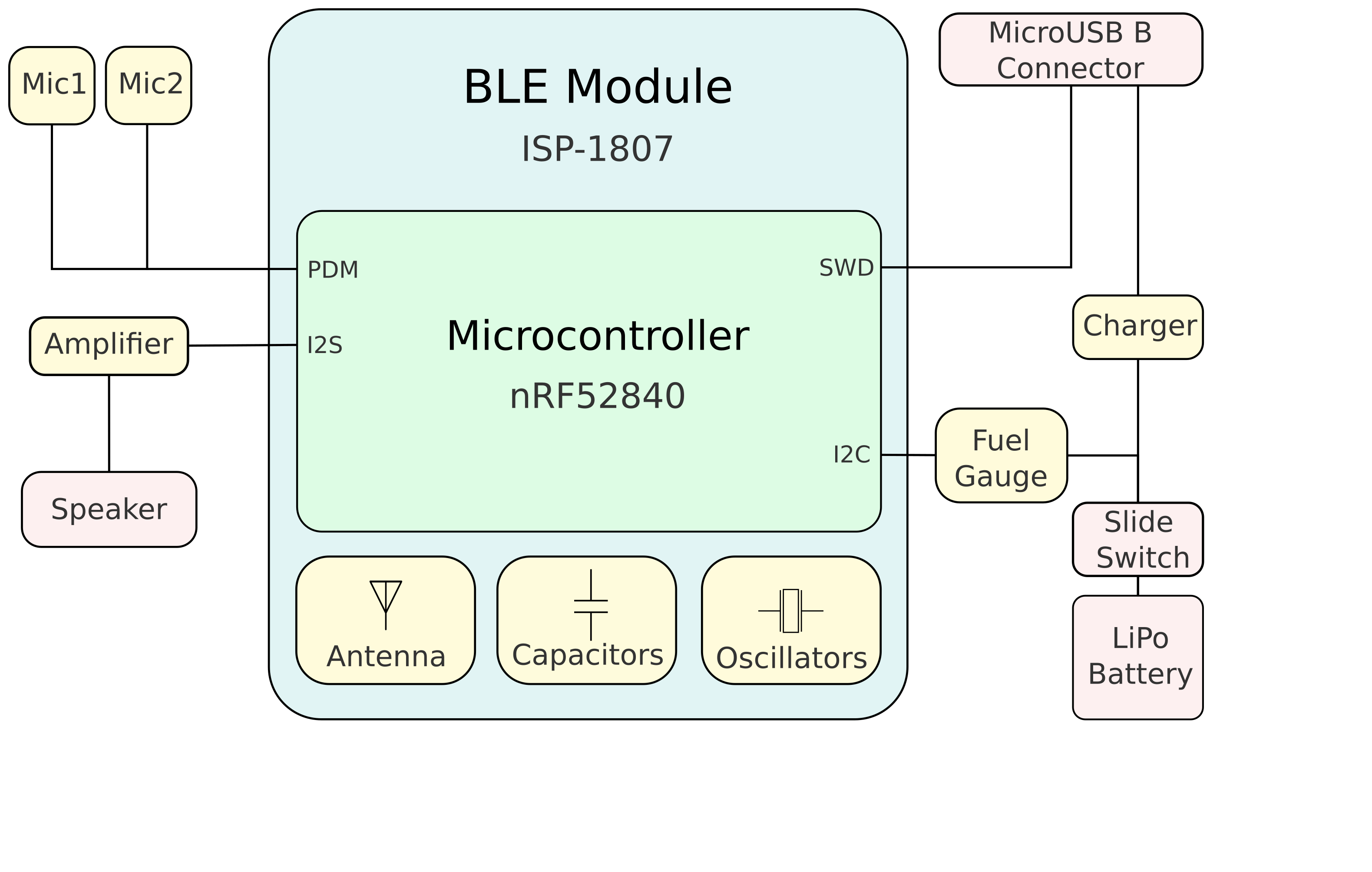}
    \vskip -0.1in
\caption{{\bf OAEbuds hardware design.} We include an additional microphone for future research.}
\vskip -0.15in
\label{fig:oaebuds_hardware}
\end{figure}

We design a custom hardware solution based on the ISP1807 Bluetooth Low Energy (BLE) module, which combines a Nordic nRF52840 microcontroller with a variety of other components such as capacitors, oscillators and an antenna. The device is equipped with a pair of pulse-density modulated (PDM) microphones (TDK Invensense T3903) and a speaker (PUI Audio AS01008MR-3) driven by a digital pulse-code modulation (PCM) input Class D amplifier (Maxim Integrated MAX98357A). The system is powered by a 3.7~V, 100~mAh Lithium Polymer Battery, and a buck converter (Texas Instruments LM3671) is used to bring the system voltage down to 3.3V. A Micro-USB connector is used to program the device over SWD and charge the battery via a charger IC (Analog Devices LTC4124). Battery information, such as cell voltage and state of charge (SOC), is probed using a fuel gauge (Maxim Integrated MAX17048). A high level overview of the system is shown in Fig.~\ref{fig:oaebuds_hardware}.

The speaker amplifier is interfaced using the controller's Inter-IC Sound (I2S) module. The device is preloaded with a fixed array in the controller's memory that holds the PCM representation of a signal (e.g., pulse). The preloaded waveform must be generated at a sampling frequency matching that of the I2S module clock signals. The sampling frequency is set to 15.625~kHz, the smallest frequency compatible with our amplifier that is larger than twice the pulse bandwidth. When the device starts emitting pulses, the controller supplies a copy of the signal waveform to the I2S module, which transfers the waveform to the amplifier using its direct memory access (DMA). The I2S module is internally double buffered, and it triggers a callback for the controller to supply a fresh audio buffer to output once a previous transfer finishes. %The controller keeps track of how many transfers have taken place, and stops the I2S module once <<X>> transfers have occured. This causes the speaker to output the preloaded waveform <<X>> times.

The microphones are interfaced using the controller's PDM module. Specifically, the module's DMA is used to asynchronously convert microphone PDM measurements to PCM values and load the results into memory. The two PDM microphones are connected to the same serial clock line, running at 1~MHz, which the PDM module internally decimates by a factor of 64. This yields an overall sampling frequency of 15.625~kHz. The recording process also uses double buffering and produces 312 channel-interleaved pairs of 16-bit samples (equivalent to 20~ms at 15.625~kHz) at a time. The captured samples are then divided into packets and transmitted over Bluetooth. To maximize throughput, each packet contains 80 two-channel samples (240 bytes total), which, when including the sequence number and overhead bytes, is the largest number of samples we can transmit in a single packet. Additionally, we also use the maximum possible data rate of 2~Mbps.  In our design, the earbuds stream the recorded  acoustic signals via Bluetooth to a nearby smartphone and the computation to extract the OAEs is performed on the smartphone.

\begin{table}[t!]
\begin{center}
  { \centering
    \footnotesize
    \begin{tabular}{|c|c|}
        \hline
        \textbf{Component} & \textbf{Cost (USD)}\\
        \hline
        BLE Module & 10.67 \\
        Microphones & 2 $\times$ 0.80 \\
        Speaker & 1.06 \\
        Amplifier & 1.62 \\
        Charger & 5.12 \\
        Fuel Gauge & 1.89 \\
        MicroUSB Connector & 0.29 \\
        Switch & 1.52 \\
        Battery & 1.50 \\
        PCB Fabrication \& Assembly & 1.63 \\
        3D Printed Case & 1.40 \\
        \hline
        Total & 28.30 \\
        \hline
    \end{tabular}
    }
    \caption{\textbf{Itemized hardware cost.} Component prices are estimated for a production lot of 1,000 devices.}
    \vskip -0.15in
    \label{tab:table1}
    \end{center}
\end{table}

\begin{figure*}[t!]
    \centering
    \begin{subfigure}[b]{.32\textwidth}
      \includegraphics[width=\textwidth]{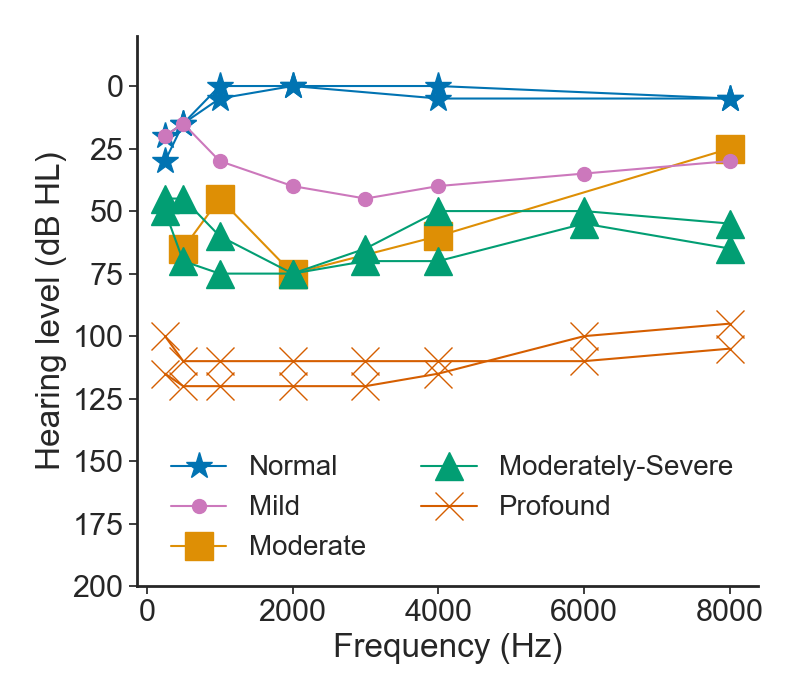}
\vskip -0.1in
      \caption{}
    \end{subfigure}
    \begin{subfigure}[b]{.32\textwidth}
      \includegraphics[width=\textwidth]{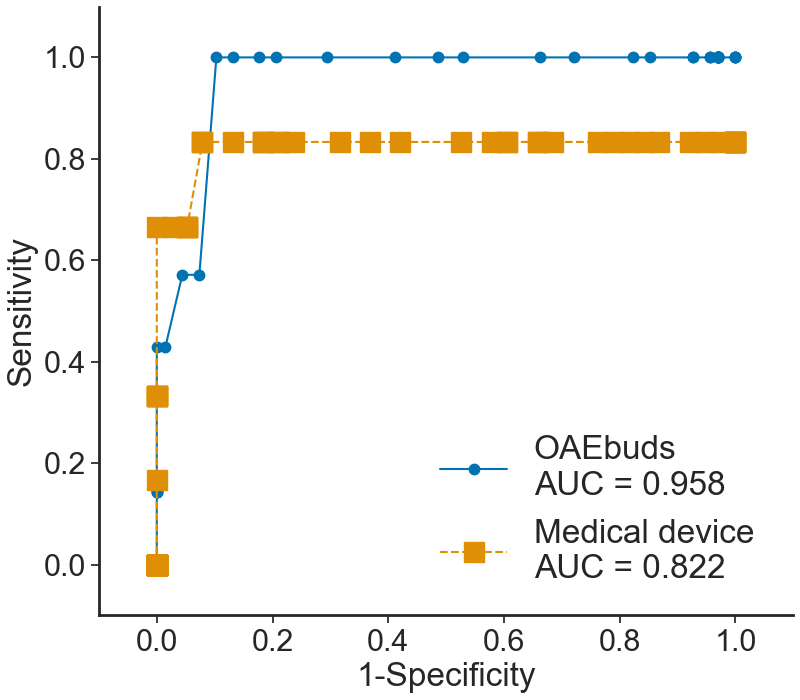}
      \vskip -0.1in
      \caption{}
    \end{subfigure}
    \begin{subfigure}[b]{.32\textwidth}
      \includegraphics[width=\textwidth]{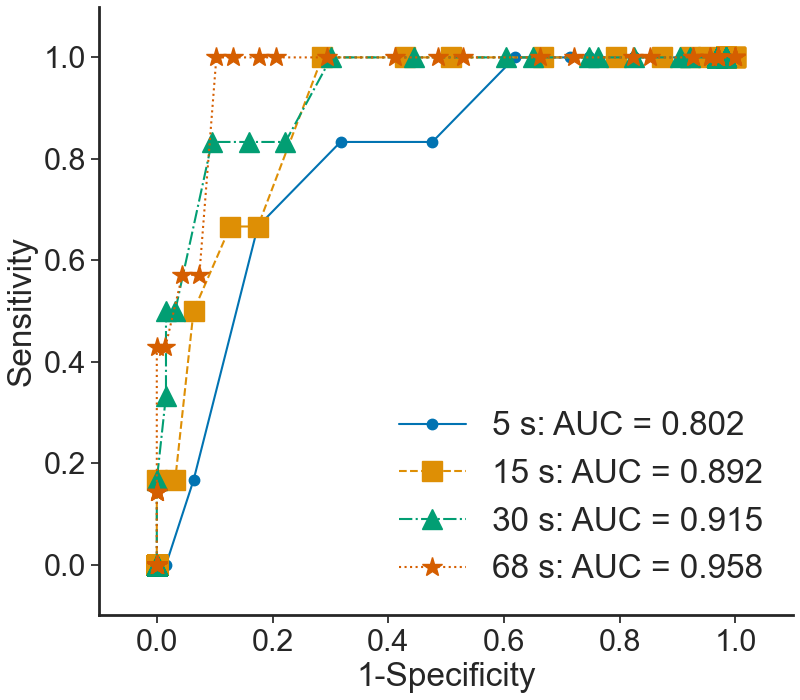}
      \vskip -0.1in
      \caption{}
    \end{subfigure}
    \vskip -0.1in
\caption{{\bf Clinical study performance.}(a) Audiograms for ears tested in clinical study with normal hearing and different degrees of hearing loss. (b) Performance of OAEbuds in comparison with commercial OAE medical device. (c) Effect of measurement time on clinical performance. }
\vskip -0.15in
\label{fig:comparison}
\end{figure*}

The circuit schematic and PCB layout for the device was designed using KiCAD and was fabricated and assembled by PCBWay. The enclosure was designed in Fusion360 and 3D-printed using a Formlabs Form 3 resin printer. Our enclosure is designed to have a tip diameter and length of 5.4 mm and 7.1 mm respectively, which allows the rubber ear tips to have a snug fit with the enclosure. We note that for all the ear tips in our study, the base diameter is the same, and is able to fit easily on the enclosure. The enclosure also has openings for a switch to power on and off the device, as well as for a micro-USB charging port. The interior of the case is also lined with foam to reduce the effect of acoustic reflections from within the case. Table.~\ref{tab:table1} shows the cost of the individual components in our earbud device estimated using Digikey, Mouser, Alibaba and PCBWay. The above numbers provide a ballpark cost  which can be further reduced at higher volumes.

 %supports a wireless platform to achieve synchronization between microphones across two earbuds and is designed for binuaral speech enhancement for telephony applications. Like eSense, the Clearbuds platform  does not have speakers or microphones facing the ear canal and is not designed for hearing loss screening.

%% file: eval.tex
% Airpods technical specifications: https://www.soundguys.com/apple-airpods-pro-review-27106/

\section{Clinical study} 
\vskip 0.15in 
Our study was approved by the Institutional Review Board and we obtained informed consent for all adults and parental consent was obtained for pediatric patients and patients aged 7 to 17 provided verbal or written assent. We recruited patients from otolaryngology, craniofacial and hearing loss clinics across  two clinical sites. We also recruited adults without any known concern for hearing loss  ($n=28$ ears). We tested our device on 50 ears from 26  newborns and  adults up to 32 years (mean age: $18 \pm 9$). We   performed testing on 10 ears between 1 week and 1 year of age. Measurements on adult patients were performed in duplicates  whenever possible, and a total of 75 measurements are used for subsequent analysis. The female-to-male ratio was 3.2.

To determine the ground truth hearing status of each patient, the best available clinical information was interpreted by the attending physician or clinician. This information includes the patient's newborn hearing screen, audiogram, diagnostic auditory brain response, and clinical and examination history. Our patient population  included  sensorineural ($n=5$) and conductive ($n=1$) hearing loss, as well as auditory neuropathy ($n=1$) which is a form of hearing loss that affects the auditory nerve's ability to transmit sound to the brain, but which does not affect the cochlea's ability to produce OAEs. We recruited patients with different degrees of hearing loss spanning the full range of degrees from slight to profound (Fig.~\ref{fig:comparison}(a)). The degree of hearing loss for a given ear is computed by taking the mean hearing level measured from a patient's audiogram in dB HL, and mapping it to the thresholds as defined by the American Speech-Language-Hearing Association~\cite{asha}. Our dataset also had ears with  middle ear infections due to fluid buildup ($n=3$), as well as ears that recently had ear tubes ($n=2$). In total, 6 ears were classified as having hearing loss, the remaining 44 ears were classified as having normal hearing or having healthy outer hair cells in the cochlea. For our study, we mark the patient with auditory neuropathy as having healthy outer hair cells in the cochlea, as OAEs are expected in this patient.

During testing, all participants $>$ 6 months  were instructed to sit upright for the test. Younger patients were tested in the position that was most comfortable for them and their parents, and included being asleep in a supine position, or being cradled over the parent's shoulder. All patients were first tested with the commercial OAE device  in each ear. Patients recruited from one of the sites were tested with a commercial TEOAE device (Otoport Screener, Otodynamics) that was used regularly at the clinic. This test was performed  across the 1, 1.5, 2, 3, 4~kHz bands. We set the device to continue measuring for this full duration even if the test passed early. The remaining patients were measured using a DPOAE device (OAE Hearing Screener, Welch Allyn) that was available for us to use at other locations.  This device tested at frequency bands of 2, 3, 4, and 5~kHz. During this portion of the test, we would select a disposable rubber ear tip (Grason \& Associates LLC) size based on visual examination of the patient's ear canal. In our study, ear tip sizes 8, 9, 10, 11 an 12 were used 9, 4, 23, 8 and 2 times respectively. For the pediatric population, we used three different ear tip sizes from 8 to 10~mm, while for the adult population four different ear tip sizes from 8 to 12~mm. This suggests that a relatively small number of ear tip sizes can accommodate a large range of ear canal sizes. We note that commercial earbuds such as AirPods contain four different ear tip sizes~\cite{airpod_eartip}.

%If the probe check on the commercial device did not detect a sufficient seal with the patient's ear, we would select a larger tip size in increments of 1~mm until this probe check passed. If the probe check did not pass after several attempts, we would terminate measurement with the commercial device.

% {\color{red}{Correct the 60 and 90 second number to put in the effective numbers assuming you are also not sending the -3 click.}}

\begin{figure*}[t!]
    \begin{subfigure}[b]{.32\textwidth}
      \includegraphics[width=\textwidth]{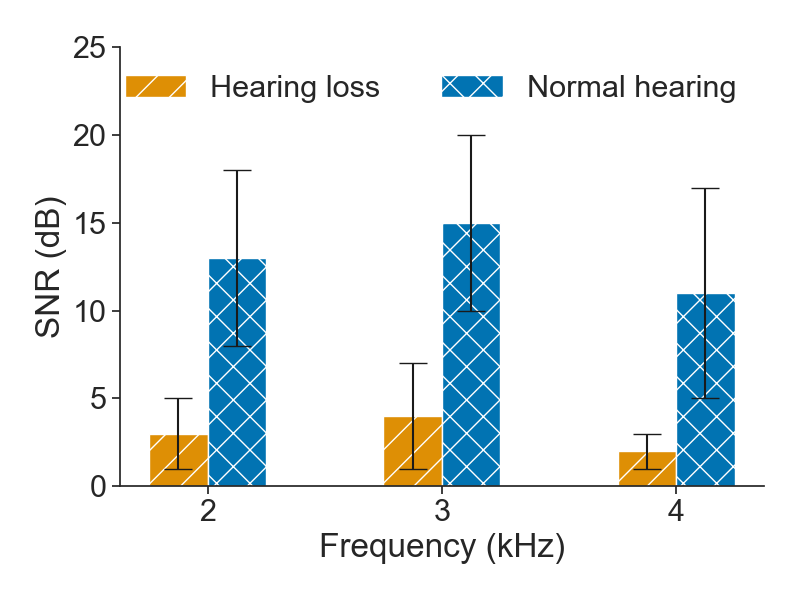}
      \caption{}
    \end{subfigure}
    \begin{subfigure}[b]{.32\textwidth}
      \includegraphics[width=\textwidth]{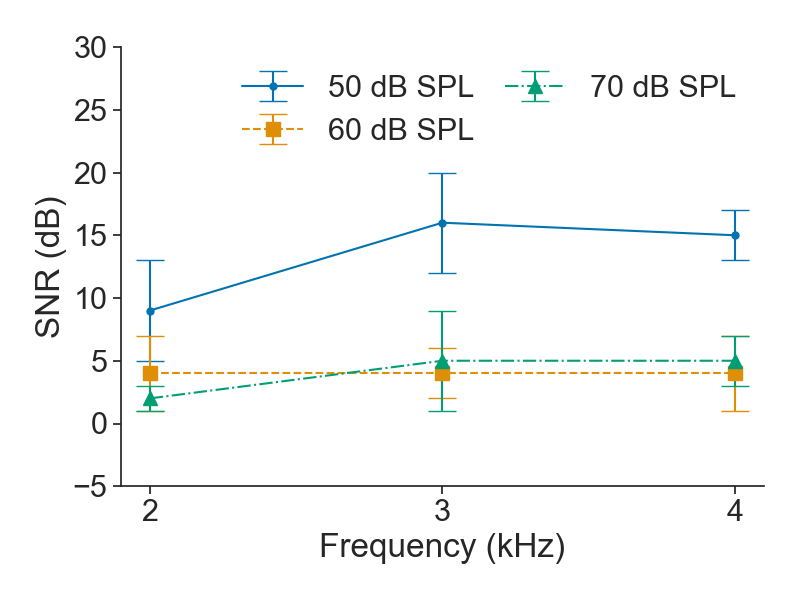}
      \caption{}
    \end{subfigure}
    \begin{subfigure}[b]{.32\textwidth}
      \includegraphics[width=\textwidth]{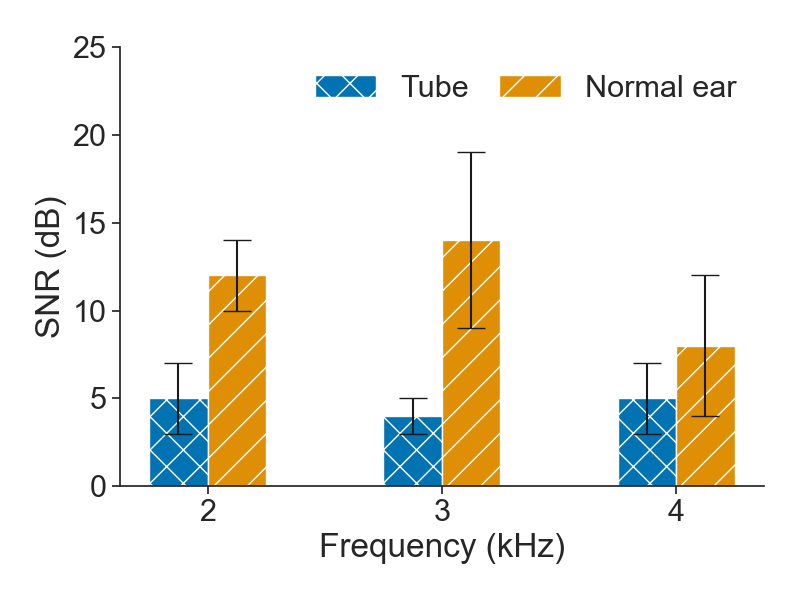}
      \caption{}
    \end{subfigure}
    \vskip -0.1in
\caption{{\bf Subgroup analysis and benchmark results.} (a) Subgroup analysis comparing the mean SNR of OAEs measured in patients with hearing loss and normal hearing during the clinical study. (b) Effect of background noise on system performance for different sound levels. (c) Probe integrity check in a close-ended tube is used to ensure that the system produces SNRs below the cutoff for healthy hearing when measured outside the ear.}
\vskip -0.15in
\label{fig:benchmark}
\end{figure*}

After completing the test with the commercial device, we proceeded to test the patient with our wireless earbud device using the same rubber ear tip. For pediatric patients, each ear was tested effectively for 45 to 68 seconds per ear, depending on the compliance of each patient. Adult participants  were tested for 68 seconds in each ear, twice. All testing with both children and adults was performed by two computer science graduate students. During testing, we transmitted clicks with a duration of 500~$\mu$s and gaps of 20~ms between clicks. The clicks were set to have a bandwidth of 0 to 5~kHz. The clicks were sent at a sound level of 84~dB peSPL (pe = peak-equivalent). % using a reference sound level meter (Amprobe SM-10). We note that the unit of dB peSPL is often used to measure the sound level of short clicks or impulses~\cite{laukli2015calibration}. This is because conventional sound level meters typically produce sound level estimates over windows ranging from 125~ms to 1~s~\cite{amprobe_manual}, which would underestimate the sound level of short clicks. We calibrate the sound level of the clicks at the earbud microphone so that the recorded peak to peak digital value is equivalent to the peak to peak value of a 1~kHz sine wave with a sound level of 84~dB SPL as measured by a sound level meter. 

 The sampling rate of the speaker and microphone was set to 15625~Hz to allow for streaming the data  over Bluetooth. We measure for OAEs at the 1, 1.5, 2, 3, and 4~kHz bands. Specifically, we average the signal and noise  responses in the ranges of 750--1250~Hz, 1250--1750~Hz, 1750--2500~Hz, 2500--3500~Hz, and 3500--4500~Hz. On our device, we consider a measurement a passing screen if the SNR of at least two frequency bands exceeds an SNR threshold of 8~dB, and the absolute sound level of those passing frequency bands is greater than -10~dB SPL, which prior work regarded as the minimum power for  these OAEs~\cite{teoae_protocol}. %This ensures that environmental noise or device artifacts are not incorrectly considered as an OAE response. In subsequent analysis, we include data from four ears from an initial wired prototype connected to a smartphone to conduct TEOAEs, and pool this data for the dual and single mic analysis.

\subsection{Performance evaluation}
\vskip 0.15in
To determine our SNR threshold on each frequency band, we generate a receiver-operating curve (Fig.~\ref{fig:comparison}) to compute the sensitivity and specificity values for SNR thresholds ranging from -20 to 40~dB in increments of 1~dB. We find that the SNR threshold of 8~dB maximizes the sum of sensitivity and specificity, yielding a sensitivity  of 100.0\% (95\% CI, 64.6--100.0\%) and specificity of 89.7\% (95\% CI, 80.2--94.9\%).  We find that using two frequency bands as the pass criteria yields an AUC of 0.958. Using three or four frequency bands as the pass criteria results in  AUCs of 0.950 and 0.884 respectively.

%, with the optimal SNR threshold of 6~dB
\vskip 0.05in\noindent{\bf Comparison with medical device.}
In comparison to our earbuds, Fig.~\ref{fig:comparison}(b) shows that the medical device had a lower AUC of 0.822 yielding a sensitivity of 83.3\% (95\% CI, 43.6--97.0\%)  and specificity of 92.1\% (95\% CI, 79.2--97.3\%). The ground truth for both our device and the medical device is the clinical information that is interpreted by the physician including their clinical and examination history as well as auditory brain response tests. We note that for one ear with hearing loss, the medical device was unable to pass the probe check despite numerous attempts to fit the ear with different sized ear tips, even though the ear tip appeared to fit well visually. For  this instance, we marked the medical  device as having failed the measurement. 

% Next, we show the agreement between our device and the commercial device for detecting OAEs. As the frequency range measured on the commercial DPOAE and TEOAE devices were different, we perform this analysis on the frequency range that is common between those devices at 2, 3, and 4~kHz. This test is performed on the 44 ears (69 measurements) where testing on the patient was performed with both the commercial and wireless earbud device. The AUC for these three frequencies is 0.615, 0.866, 0.681 respectively. We note that for 32 of the 44 ears, the commercial device we used was a DPOAE device. We note that the absolute sound level of DPOAEs are typically higher than TEOAEs~\cite{xxx}. In a measurement of both OAE tests on four normal hearing ears, we find that the DPOAE signal is on average 10~dB higher than the TEOAE signal across the 2, 3, and 4~kHz band.

Our device misclassified the hearing loss status for seven ears, three of these ears had middle ear fluid and infection. In these ears, OAEs were not detected, as the fluid acts as a barrier that blocks the OAEs from reaching the outer ear, and it is expected that OAEs do not appear in these ears~\cite{boone2005failed}. The commercial device similarly did not detect OAEs in ears with middle ear fluid or infection. One of these ears had hearing loss, but OAEs were detected. We suspect that this is due to reflections or ear tip fit issues, as this was the only ear where the commercial device could not begin a test due to a failure of the initial probe check despite repeated attempts to pass the check. One of these ears recently had ear tubes removed which would have resulted in a hole in the eardrum which would have begun healing, and may have affected the ability to detect OAEs.

% Although the measurement time of our device was set to 90 seconds to transmit the sequence of $\{1,1,1,-3\}$ clicks, the effective measurement time using our protocol is 67.5 seconds (3294 clicks for one mic) as we only perform analysis on the clicks with positive polarity to compute TEOAEs on our system. This measurement time was set conservatively in order to collect as much data as possible, 
% we can obtain good clinical performance even when the measurement time is reduced to 15 or 30 seconds (1462 and 2926 clicks per mic respectively) 

\vskip 0.05in\noindent {\bf Effect of measurement time.} 
To determine the effect of measurement time, we set a limit on the maximum number of clicks used by our algorithm and compute clinical performance when the measurement time is reduced to 15 or 30 seconds. Fig.~\ref{fig:comparison}(c)  shows that we are able to achieve  an AUC of 0.892 and 0.915, which is higher than that achieved by the commercial device of 0.822. At these measurement times, our system obtained an optimal sensitivity of 100.0\% (95\% CI, 61.0--100.0\%) and 83.3\% (95\% CI, 43.6--97.0\%) respectively and specificity of 71.4\% (95\% CI, 59.3--81.1\%) and 90.5\% (95\% CI, 80.7--95.6\%) respectively.  When reducing the measurement duration to 5 seconds, our sensitivity and specificity, as expected, reduces to 83.3\% (95\% CI, 43.6--97.0\%) and 68.3\% (95\% CI, 56.0--78.4\%) respectively.

\vskip 0.05in\noindent {\bf Subgroup analysis of hearing status.}
In Fig.~\ref{fig:benchmark}(a), we show the average SNR obtained  in ears with hearing loss as well as with normal hearing. The plot shows that for the ears with hearing loss the average SNR across all frequencies is 3~dB, while it is 11~dB for ears with normal hearing, showing that there is large separation in SNRs between the two classes of ears. We note that in the hearing loss ears, although the average SNRs are positive, none of the SNRs at any of the frequencies exceeded the 8~dB SNR cutoff. We suspect the SNRs are positive due to residual reflections from within the plastic case of our wireless earbuds. We also note that the SNRs for the OAEs detected by our system are smaller at the lower frequencies. This is in line with existing literature~\cite{debra_teoae_1998} which confirms that transient-evoked OAEs are better at mid-range frequencies than the lower frequencies.

%We note that for both the commercial device and our wireless earbuds, the SNRs are generally higher in the 2--4~kHz range. 

\vskip 0.05in\noindent {\bf Test-retest evaluation.} 
For the 25 ears where duplicate testing was performed, the screening result for our earbuds  matched in all but one ear. In our study, we also tested the commercial device several times if we were not confident in the probe fit in the ear during a given measurement. There were two ears where the commercial device had differing results between tests. Both of these ears were normal hearing ears and it took two and three attempts respectively for these ears in order to obtain a passing screen result.

\section{Micro-benchmarks}\label{sec:benchmark}
\vskip 0.15in
We provide various micro-benchmark evaluations including the effect of background noise and a comparison of our wireless sensing techniques with existing OAE algorithms. We also present an evaluation of system level issues including power and run-time analyses. 
\vskip 0.05in\noindent{\bf Effect of background noise. }To measure the effect of background noise on our device, we played noise of road traffic from a laptop such that the sound level at a healthy ear varied from 50 to 70~dB SPL and measured the OAEs measured by our earbuds in an ear with  normal hearing. These sound levels reflect the typical range of ambient background noise that would occur in a clinical testing facility that is not well insulated from noise, and which might be situated close to a road. Fig.~\ref{fig:benchmark}(b) shows that at a noise of 50~dB  SPL, the OAEs can be detected in the ear, and are above the 8~dB threshold at all frequencies. This sound level is the typical ambient noise level in an urban residence~\cite{King2012-ah}. At 60 and 70~dB SPL, the SNRs at all frequencies drop below the 8~dB threshold. These sound levels are similar to conversational speech held at 1~m and a vacuum cleaner at 1~m~\cite{common_spl_levels}. These results suggest that testing should be performed in a relatively quiet environment to obtain reliable results.

\vskip 0.05in\noindent {\bf Probe integrity check.} Medical OAE devices use a closed-ended tube between 0.5--2~cc in volume as a probe integrity check~\cite{teoae_protocol}. This range of volumes is selected to mimic the volume of the ear canal for the pediatric and adult population.
Similarly, we have our earbuds  perform a measurement in a closed ended plastic tube  with a volume of 1~cc, and in a healthy ear. Fig.~\ref{fig:benchmark}(c)  shows the SNRs obtained in both these scenarios when repeated three times. The plot shows that the SNRs in the tube are below the SNR cutoff of 8~dB for all frequencies across all measurements, and can be used as a probe integrity check to ensure that the device is not incorrectly identifying OAEs. 

%We additionally shows the performance of our device in open air. We notice that in this scenario, there are significant number of reflections which produce a high SNR, and so a measurement in open air cannot be reliably used for a probe integrity check.

\begin{figure}[t!]
\centering
        \begin{subfigure}[b]{.35\textwidth}
      \includegraphics[width=\textwidth]{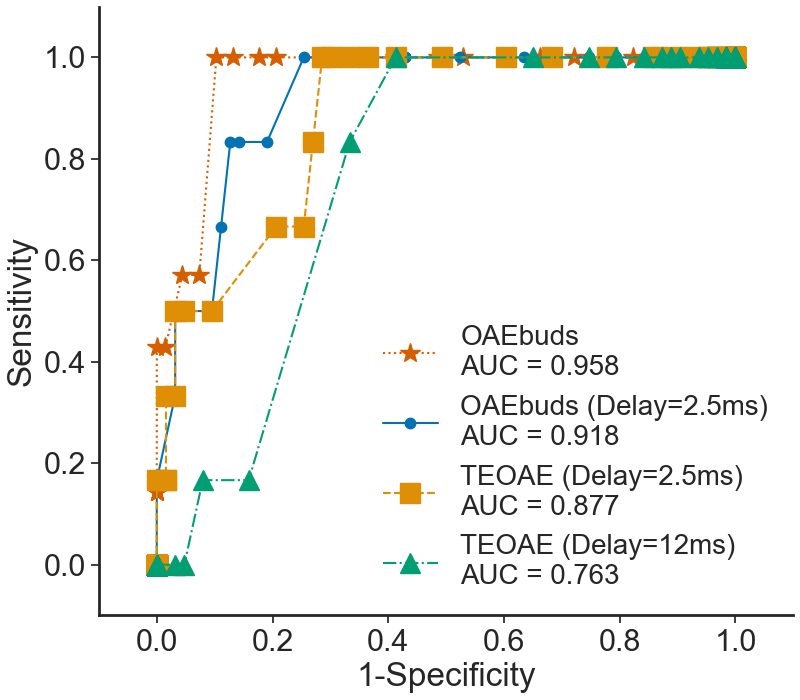}
      \vskip -0.06in
      \caption{}
    \end{subfigure}
        \begin{subfigure}[b]{.35\textwidth}
      \includegraphics[width=\textwidth]{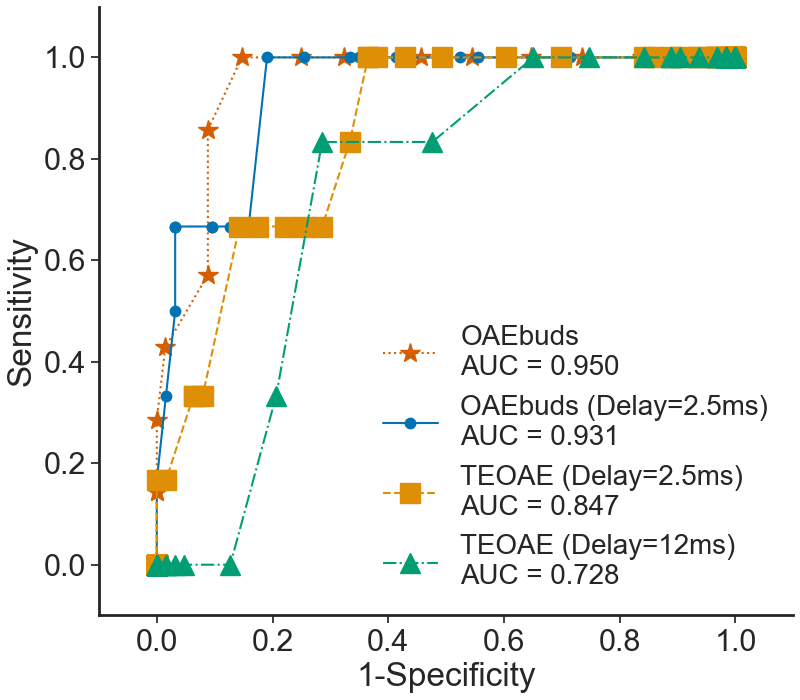}
      \vskip -0.06in
      \caption{}
    \end{subfigure}
    \vskip -0.1in
\caption{{\bf Comparison of OAEbuds with prior TEOAE algorithms.} Our OAEbuds system achieves better performance compared to prior TEOAE algorithms when (a) 2 and (b) 3  frequency bands are required to be above the SNR threshold for the hearing test to pass.}
\vskip -0.15in
\label{fig:protocols}
\end{figure}

%%We transmit a train of pulses with a polarity pattern of $\{1,1,1,-3\}$. This polarity pattern was chosen to allow for analysis using both the conventional TEOAE protocol, as well as our modified protocol customized for our hardware setup.
\vskip 0.05in \noindent {\bf Comparison with prior OAE algorithms.} We evaluate the performance of our OAEbuds system using alternative variations to implementation. We test three different protocols as follows 1) OAEbuds protocol with a fixed delay of 2.5~ms where reflections after  a 2.5~ms duration from the pulse are removed, 2) conventional TEOAE protocol with a delay of 2.5~ms and 3) conventional TEOAE protocol with a higher delay of 12~ms.
Conventional TEOAE systems transmit a train of pulses with a polarity pattern of $\{1,1,1,-3\}$~\cite{basics_oae}. The receiver then adds up the response across the four pulses to generate the OAE signals. We select 2.5~ms as the delay as this is what commercial TEOAE devices typically use~\cite{teoae1}. Fig.~\ref{fig:protocols} shows the ROC curves for these protocol implementations when using either 2 or 3 frequency bands to pass. We note that our implementation of OAEbuds  yields a better AUC compared to the conventional protocol regardless of whether 2 or 3 frequency bands are needed to pass. We note that when using a delay of 2.5~ms, a significant amount of the signal will be reflections and not OAEs. Because of this, the optimal SNR threshold for these modified protocol is significantly higher at 51 and 48~dB for the 2 and 3 frequency band scenarios respectively.  The most likely  reason why the TEOAE protocol performs worse than  OAEbuds is that it relies on the clicks canceling out each other well. In practice, the cancellation is not perfect and there will be a residual error in the cancellation process, in particular on our low-cost acoustic hardware that has non-linearities. Although the goal of this protocol is to cancel out reflections which are linear, it will also cancel out the linear components of the OAEs themselves, which may also contribute to lowered performance. We note that with TEOAE, using the lower delay of 2.5~ms performs better than the 12~ms delay. With this lower delay, the optimal SNR threshold is 21--22~dB, while it is 4--6~dB when running the TEOAE protocol with a 12~ms delay.

\begin{figure}[t!]
    \begin{subfigure}[b]{.4\textwidth}
    \centering
    \includegraphics[width=\textwidth]{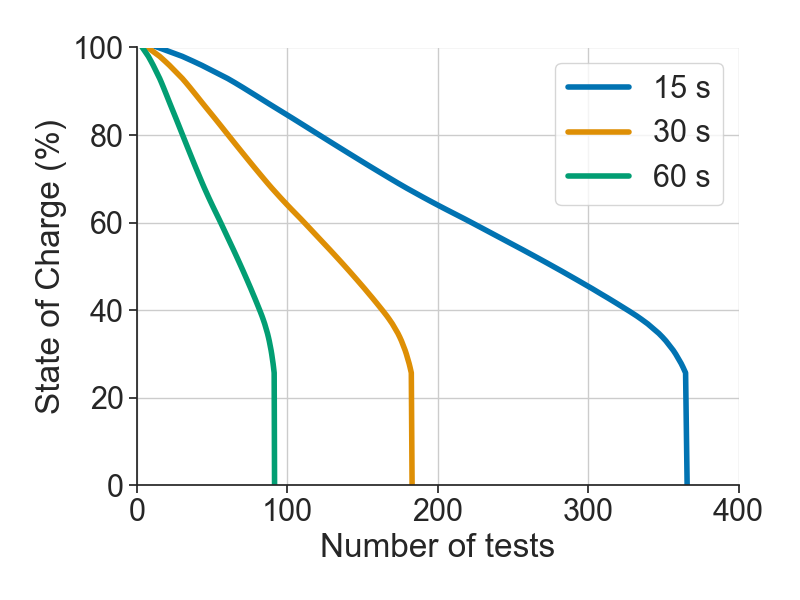}
    \vskip -0.1in
    \caption{}
    \end{subfigure}
    \begin{subfigure}[b]{.4\textwidth}
    \centering
    \includegraphics[width=\textwidth]{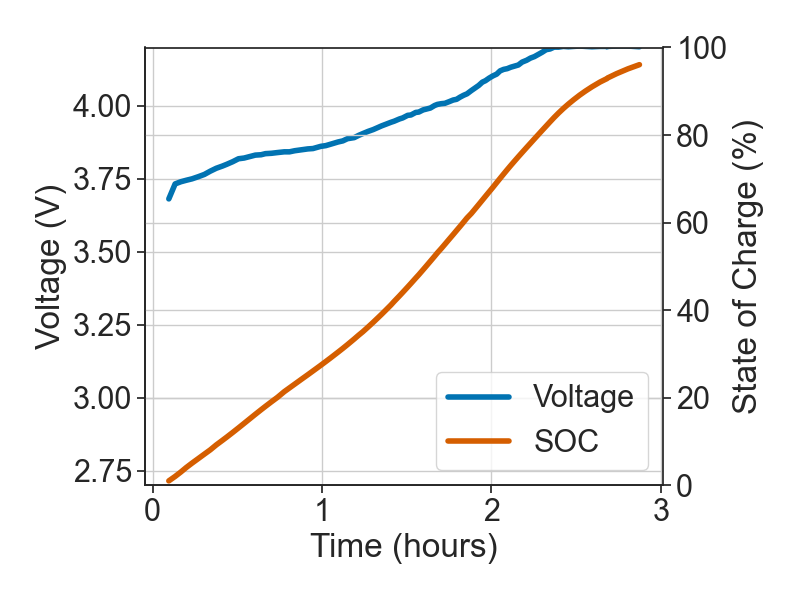}
    \vskip -0.1in
    \caption{}
    \end{subfigure}
    \vskip -0.05in
\caption{{\bf Power analysis.} (a) Number of tests that can be performed on a single charge for tests of different durations. (b) Time required to charge an OAEbud via a micro-USB connection.}
\vskip -0.15in
\label{fig:power}
\end{figure}

\vskip 0.05in\noindent {\bf Power analysis.} To evaluate how long our earbuds last on a single charge, we first charge the battery of the earbuds to its maximum level and evaluate its performance over multiple tests. Fig.~\ref{fig:power}(a) shows the state of charge on the battery as a function of the number of tests. Each of the lines represents a different duration for a hearing test. To measure the voltages and state of charge, we use an on-device fuel gauge. The fuel gauge uses the proprietary ModelGauge algorithm to continuously track the battery’s state of charge (SOC). It simulates the internal nonlinear dynamics of the battery model. By analyzing voltage measurements over time, it can determine the state of charge much more accurately. The fuel gauge data is interfaced over an I2C bus and it allows us to read out the battery voltage and SOC.

The plots show that on a single charge, even when each test lasts around 60~s, the earbud can support 91 hearing tests. For context, we note that on a typical day, the hearing loss clinic in our institution sees around 20-30 patients. We measure how long it takes to charge the earbuds through a micro-USB cable connected to a wall outlet. Fig.~\ref{fig:power}(b) shows that it takes around 3 hours to charge the earbud which in a practical setting can be done overnight in a clinic. 

\vskip 0.05in\noindent {\bf Runtime analysis.} To evaluate the feasibility of running our  algorithm on a mobile device like a smartphone, we convert our algorithm to C++ code that can run on the Android smartphone platform and time how long it takes to execute the operations in our system. On a Samsung Galaxy S9, we are able to process windows of 1~s containing 48 clicks in less than a millisecond. This means that we are able to provide a real-time update of the SNR values throughout a OAE measurement. This real-time feedback can help a user determine if a probe fit is snug or if there is too much noise at different points in a measurement, and allows them to make any changes to how the measurement is being performed. 

%% file: related-1.tex
\section{Related work}
\vskip 0.15in

%In the rest of section, we describe other related work in the broad space of earables.

Prior work can be broadly divided into three classes.

\vskip 0.05in\noindent{\bf Health tracking using earphones.} Prior work has explored the use of earphones for monitoring physiological signals  for cardiovascular sensing~\cite{cardiac1,headfi}, blood pressure measurements~\cite{ebp} and  respiration~\cite{respiratoryrate}. Earphones have also been used for sensing  jaw clenching~\cite{jawclenching}, teeth motion  and  voice detection~\cite{teethactivity,voiceactivity}. 
Prior work has  also explored the use of smartphone attachments for diagnosing middle ear conditions. \cite{chan2019detecting} designed a  paper cone that is attached to a smartphone and used a machine learning classifier to detect  middle ear fluid behind the ear drum. \cite{earhealth} used a smartphone-connected wired earphone attached to an external microphone  to  differentiate between   middle ear fluid, ear drum ruptures and  wax blockage. \cite{chan_justin_2022_6389337} presented a smartphone attachment to perform tympanometry where the pressure  within the ear canal is changed to assess the mobility of the ear drum. All these  wired systems however are designed for assessing the state of the middle ear and the ear drum. Hearing screening, in contrast, is primarily focused on the state of the cochlea. The cochlea is part of the inner-ear that is behind the ear drum and is primarily responsible for converting sound waves into electrical impulses which are then interpreted by the brain.

\vskip 0.05in\noindent{\bf  OAE  devices.}  Prior work  \cite{heitmann2021mass,heitmann2017sound4all} created a smartphone interface for the probes from an existing commercial OAE device \cite{pathmedical}. In addition to not being wireless, this approach is still constrained by the cost of commercial OAE probes that are expensive. Further, since it is directly  connected to a smartphone it requires calibration for each smartphone model which is difficult to generalize.    Recent commercial approaches~\cite{otoscreen1,otoscreen2}  use  bone conduction to monitor OAEs through a headband.  In addition to not being in the wireless earbud form factor, these have not been demonstrated to be low-cost.   \cite{heitmann2020designing} proposes to use a single transducer to measure OAEs. However it is primarily focused on the transducer characterization and does not build an end-to-end wireless earbud system. Further, none of these prior efforts have performed clinical studies  to evaluate  efficacy.  Finally, high-end personalized  earphones from companies like Nura Sound  use   OAE measurements to customize music for an  adult wearer~\cite{nurapatent,nura}. Our goal in this work is complementary in that we  create a low-cost and open-source  earbud system that is designed to achieve   hearing loss screening with accuracies similar to medical devices. %Further, our system is designed to  work across a diverse age range from new-born babies to adults.

\vskip 0.05in\noindent{\bf Earable platforms.} Recent years have seen the introduction of earbud platforms like eSense platform~\cite{esense1,esense2},  Clearbuds~\cite{clearbuds}  and OpenEarable~\cite{openearable}.  Like eSense, the Clearbuds platform  does not have speakers or microphones facing the ear canal and is not designed for hearing loss screening.  OpenEarable  has a rigid over the ear design that is hard to operate across different age groups. We instead design an open-source wireless earbud platform  to reliably measure otoacoustic emissions for hearing loss screening using low-cost hardware.

 %supports a wireless platform to achieve synchronization between microphones across two earbuds and is designed for binuaral speech enhancement for telephony applications. Like eSense, the Clearbuds platform  does not have speakers or microphones facing the ear canal and is not designed for hearing loss screening.

%% file: limits.tex
\section{Limitations and discussion}
\vskip 0.15in
We describe various limitations of our current system and discuss the regulatory pathway.

\vskip 0.05in\noindent{\bf Field studies.} In practical deployments,  OAEbuds will potentially be used by a range of stakeholders including nurses, technicians and volunteers. Our clinical study does show that graduate students with no formal training in audiology were able to select the ear tips and snugly place the earbuds for both infants and adults. However,  field studies  in low and middle-income countries might be required to ensure that our design can be used as advertised; this is however not in the scope of this paper.

\vskip 0.05in\noindent{\bf Followup care and regulatory costs.} Detecting hearing loss is an important first step in addressing this complex public health problem. Other factors include  human resources for performing the tests, followup care and regulatory costs. We however note that prior FDA clearances for OAE devices did not require  human testing~\cite{fda}, which significantly reduces the cost of clearance. 

\vskip 0.05in\noindent{\bf Software update to commercial earbuds.} We  develop a custom  earbud instead of using existing  earbuds for two  key reasons: 1) we wanted to achieve a lower cost than existing wireless earbuds and demonstrate that OAEs can be detected using low-cost acoustic components, and 2) commercial wireless earbuds do not provide  access to data from the in-ear microphone. We however note that the microphones and speaker used in  Apple AirPods and Pixel buds have a higher quality. Given that all the hardware including an in-ear microphone are already present in commercial earbuds, our paper shows that there is an exciting possibility that using the algorithms presented here, commercial earbuds can potentially enable OAE detection and hearing loss screening using only a software  update.

%% file: conclude.tex
\section{Conclusion}
\vskip 0.15in
Over the next decade, the mobile systems community is uniquely positioned  to  develop wearable and mobile technologies that  help alleviate global health inequity. We developed the first wireless earbuds that can detect otoacoustic emissions and perform hearing screening using low-cost acoustic hardware. Our work introduces two  components, 1) a low-cost wireless earbud hardware  for hearing screening that  works across infants and adults, and 2)  wireless sensing algorithms to reliably identify otoacoustic emissions in the presence of in-ear reflections and echoes. Our clinical study  demonstrates similar sensitivity and specificity to commercial medical devices and shows the potential of our design to enable hearing loss screening in low and middle income countries.

%% file: ms.bbl
\begin{thebibliography}{10}

\bibitem{who_hl}
Millions of people in the world have hearing loss that can be treated or
  prevented.
\newblock {\em World Health Organization}, 2013.

\bibitem{mcpherson2012newborn}
Bradley McPherson.
\newblock Newborn hearing screening in developing countries: Needs \& new
  directions.
\newblock {\em The Indian {J}ournal of {M}edical {R}esearch}, 135(2):152, 2012.

\bibitem{swanepoel2010telehealth}
De~Wet Swanepoel, Jackie~L Clark, Dirk Koekemoer, James~W Hall~III, Mark Krumm,
  Deborah~V Ferrari, Bradley McPherson, Bolajoko~O Olusanya, Maurice Mars,
  I{\^e}da Russo, et~al.
\newblock Telehealth in audiology: The need and potential to reach underserved
  communities.
\newblock {\em International Journal of Audiology}, 49(3):195--202, 2010.

\bibitem{standard2}
Joint (JCIH.
\newblock Joint committee on infant hearing year 2000 position statement:
  Principles and guidelines for early hearing detection and intervention.
\newblock {\em Pediatrics}, 106:798--817, 01 2000.

\bibitem{standard3}
Stephen Epstein, Terese Finitzo, Allen Erenberg, and Nancy Roizen.
\newblock Joint committee on infant hearing 1994 position statement.
\newblock {\em ASHA}, 5, 01 1991.

\bibitem{nbme22}
Justin Chan, Nada Ali, Ali Najafi, Anna Meehan, Lisa Mancl, Emily Gallagher,
  Randall Bly, and Shyamnath Gollakota.
\newblock An off-the-shelf otoacoustic-emission probe for hearing screening via
  a smartphone.
\newblock {\em Nature Biomedical Engineering}, 6:1--11, 10 2022.

\bibitem{standard1}
Universal newborn hearing screening.
\newblock {\em American family physician}, 75:1349--52, 06 2007.

\bibitem{standard4}
U.{S}. {D}epartment of {H}ealth and {H}uman {S}ervices. {H}ealthy {P}eople
  2010. {W}ashington, {D.C.: U.S. D}epartment of {H}ealth and {H}uman
  {S}ervices, 2000:3–22.

\bibitem{patel2011universal}
H~Patel and M~Feldman.
\newblock Universal newborn hearing screening.
\newblock {\em Paediatrics \& {C}hild {H}ealth}, 16(5):301--305, 2011.

\bibitem{cdc}
Why is a hearing screening important for my baby?

\bibitem{app3}
Hearing {T}est \& {E}ar {A}ge {T}est, 2022.
\newblock
  \url{https://apps.apple.com/us/app/hearing-test-ear-age-test/id1067630100}.

\bibitem{app1}
Mimi hearing test, 2022.
\newblock \url{https://apps.apple.com/us/app/mimi-hearing-test/id932496645}.

\bibitem{app2}
Hearing test - {A}udiometry {T}one, 2022.
\newblock
  \url{https://apps.apple.com/us/app/hearing-test-audiometry-tone/id1368396053}.

\bibitem{walker2013audiometry}
Jennifer~Junnila Walker, Leanne~M Cleveland, Jenny~L Davis, and Jennifer~S
  Seales.
\newblock Audiometry screening and interpretation.
\newblock {\em American {F}amily {P}hysician}, 87(1):41--47, 2013.

\bibitem{abdala2001distortion}
Caroline Abdala and Leslie Visser-Dumont.
\newblock Distortion product otoacoustic emissions: a tool for hearing
  assessment and scientific study.
\newblock {\em The Volta Review}, 103(4):281, 2001.

\bibitem{chan2022inner}
Justin Chan and Shyamnath Gollakota.
\newblock Inner-ear cochlea testing with earphones.
\newblock In {\em Proceedings of the 20th Annual International Conference on
  Mobile Systems, Applications and Services}, pages 609--610, 2022.

\bibitem{thompson2001universal}
Diane~C Thompson, Heather McPhillips, Robert~L Davis, Tracy~A Lieu, Charles~J
  Homer, and Mark Helfand.
\newblock Universal newborn hearing screening: summary of evidence.
\newblock {\em Jama}, 286(16):2000--2010, 2001.

\bibitem{torre2003distortion}
Peter Torre~III, Karen~J Cruickshanks, David~M Nondahl, and Terry~L Wiley.
\newblock Distortion product otoacoustic emission response characteristics in
  older adults.
\newblock {\em Ear and hearing}, 24(1):20--29, 2003.

\bibitem{product1}
Welch Allyn.
\newblock Welch allyn oae hearing screener.
\newblock
  \url{https://mdmaxx.com/products/oae-hearing-screener-with-printer?currency=USD&variant=38178860990655}.

\bibitem{product2}
e3~Diagnostics.
\newblock Otoacoustic emissions (oae) machines.
\newblock
  \url{https://www.e3diagnostics.com/products/otoacoustic-emissions---oae}.

\bibitem{india1}
Suneela Garg, Ritesh Singh, and Deeksha Khurana.
\newblock Infant hearing screening in india: Current status and way forward.
\newblock {\em International Journal of Preventive Medicine}, 6:113, 11 2015.

\bibitem{kenya1}
Justin Shinn, Asitha Jayawardena, Ankita Patro, M.~Geraldine Zuniga, and James
  Netterville.
\newblock Teacher prescreening for hearing loss in the developing world.
\newblock {\em Ear, Nose \& Throat Journal}, 100:014556131988038, 10 2019.

\bibitem{asha.org}
American Speech Language~Hearing Association.
\newblock Adult hearing screening.
\newblock
  \url{https://www.asha.org/practice-portal/professional-issues/adult-hearing-screening/}.

\bibitem{speed}
Interacoustics.
\newblock {OAE}s: The difference between {TEOAE}, {DPOAE} and {SFOAE}.
\newblock \url{https://www.youtube.com/watch?v=aoIVSyRqWwk}.

\bibitem{kemp_acoustic_1986}
D~T Kemp, P~Bray, L~Alexander, and A~M Brown.
\newblock Acoustic emission cochleography–practical aspects.
\newblock {\em Scandinavian audiology.}, 25, 1986.
\newblock Place: Stockholm, Publisher: Audiological Society of Denmark,
  Finland, Iceland, Norway and Sweden.

\bibitem{neumann1994chirp}
Joachim Neumann, Stefan Uppenkamp, and Birger Kollmeier.
\newblock Chirp evoked otoacoustic emissions.
\newblock {\em Hearing research}, 79(1-2):17--25, 1994.

\bibitem{chan2000test}
Rosita~H Chan and Bradley McPherson.
\newblock Test-retest reliability of tone-burst-evoked otoacoustic emissions.
\newblock {\em Acta otolaryngologica}, 120(7):825--834, 2000.

\bibitem{zemplenyi1985optical}
Jan Zemplenyi, Samuel Gilman, and Donald Dirks.
\newblock Optical method for measurement of ear canal length.
\newblock {\em The Journal of the Acoustical Society of America},
  78(6):2146--2148, 1985.

\bibitem{asha}
Degree of hearing loss.
\newblock 2022.
\newblock \url{https://www.asha.org/public/hearing/degree-of-hearing-loss/}.

\bibitem{airpod_eartip}
Choose your {A}ir{P}ods {P}ro ear tips and use the ear tip fit test.
\newblock 2022.
\newblock \url{https://support.apple.com/en-us/HT210633}.

\bibitem{teoae_protocol}
Creating a teoae pass/refer protocol, 2022.
\newblock
  \url{https://www.interacoustics.com/oae-devices/lyra/support/creating-a-teoae-pass-refer-protocol}.

\bibitem{boone2005failed}
Ryan~T Boone, Charles~M Bower, and Patti~F Martin.
\newblock Failed newborn hearing screens as presentation for otitis media with
  effusion in the newborn population.
\newblock {\em International Journal of Pediatric Otorhinolaryngology},
  69(3):393--397, 2005.

\bibitem{debra_teoae_1998}
Debra~M. Hussain, Michael~P. Gorga, Stephen~T. Neely, Douglas~H. Keefe, and
  Jo~Peters.
\newblock Transient evoked otoacoustic emissions in patients with normal
  hearing and in patients with hearing loss.
\newblock {\em Ear and Hearing}, 19(6), 1998.

\bibitem{King2012-ah}
Gavin King, Marek Roland-Mieszkowski, Timothy Jason, and Daniel~G Rainham.
\newblock Noise levels associated with urban land use.
\newblock {\em J Urban Health}, 89(6):1017--1030, December 2012.

\bibitem{common_spl_levels}
Table of sound pressure levels.
\newblock 2022.
\newblock \url{http://www.sengpielaudio.com/TableOfSoundPressureLevels.htm}.

\bibitem{basics_oae}
Basic of {OAE}s.
\newblock 2022.
\newblock \url{https://www.otoemissions.org/old/definitions/TEOAE2.html}.

\bibitem{teoae1}
{TEOAE}s {T}est {P}rocedures.
\newblock 2022.
\newblock
  \url{https://www.otoemissions.org/index.php/en/basics-of-oaes/teoaes/3-teoaes-test-procedures}.

\bibitem{cardiac1}
Ming-Zher Poh, Kyunghee Kim, Andrew Goessling, Nicholas Swenson, and Rosalind
  Picard.
\newblock Cardiovascular monitoring using earphones and a mobile device.
\newblock {\em IEEE Pervasive Computing}, 11(4):18--26, 2012.

\bibitem{headfi}
Xiaoran Fan, Longfei Shangguan, Siddharth Rupavatharam, Yanyong Zhang, Jie
  Xiong, Yunfei Ma, and Richard Howard.
\newblock Headfi: Bringing intelligence to all headphones.
\newblock In {\em Proceedings of the 27th Annual International Conference on
  Mobile Computing and Networking}, MobiCom '21, page 147–159, New York, NY,
  USA, 2021. Association for Computing Machinery.

\bibitem{ebp}
Nam Bui, Nhat Pham, Jessica~Jacqueline Barnitz, Zhanan Zou, Phuc Nguyen, Hoang
  Truong, Taeho Kim, Nicholas Farrow, Anh Nguyen, Jianliang Xiao, Robin
  Deterding, Thang Dinh, and Tam Vu.
\newblock Ebp: A wearable system for frequent and comfortable blood pressure
  monitoring from user's ear.
\newblock In {\em The 25th Annual International Conference on Mobile Computing
  and Networking}, MobiCom '19, New York, NY, USA, 2019. Association for
  Computing Machinery.

\bibitem{respiratoryrate}
Tobias R\"{o}ddiger, Daniel Wolffram, David Laubenstein, Matthias Budde, and
  Michael Beigl.
\newblock Towards respiration rate monitoring using an in-ear headphone
  inertial measurement unit.
\newblock In {\em Proceedings of the 1st International Workshop on Earable
  Computing}, EarComp'19, page 48–53, New York, NY, USA, 2020. Association
  for Computing Machinery.

\bibitem{jawclenching}
Siddharth Rupavatharam and Marco Gruteser.
\newblock Towards in-ear inertial jaw clenching detection.
\newblock In {\em Proceedings of the 1st International Workshop on Earable
  Computing}, EarComp'19, page 54–55, New York, NY, USA, 2020. Association
  for Computing Machinery.

\bibitem{teethactivity}
Jay Prakash, Zhijian Yang, Yu-Lin Wei, Haitham Hassanieh, and Romit~Roy
  Choudhury.
\newblock Earsense: Earphones as a teeth activity sensor.
\newblock In {\em Proceedings of the 26th Annual International Conference on
  Mobile Computing and Networking}, MobiCom '20, New York, NY, USA, 2020.
  Association for Computing Machinery.

\bibitem{voiceactivity}
Narimene Lezzoum, Ghyslain Gagnon, and Jérémie Voix.
\newblock Voice activity detection system for smart earphones.
\newblock {\em IEEE Transactions on Consumer Electronics}, 60(4):737--744,
  2014.

\bibitem{chan2019detecting}
Justin Chan, Sharat Raju, Rajalakshmi Nandakumar, Randall Bly, and Shyamnath
  Gollakota.
\newblock Detecting middle ear fluid using smartphones.
\newblock {\em Science Translational Medicine}, 11(492), 2019.

\bibitem{earhealth}
Yincheng Jin, Yang Gao, Xiaotao Guo, Jun Wen, Zhengxiong Li, and Zhanpeng Jin.
\newblock Earhealth: An earphone-based acoustic otoscope for detection of
  multiple ear diseases in daily life.
\newblock In {\em Proceedings of the 20th Annual International Conference on
  Mobile Systems, Applications and Services}, MobiSys '22, page 397–408, New
  York, NY, USA, 2022. Association for Computing Machinery.

\bibitem{chan_justin_2022_6389337}
Justin Chan, Ali Najafi, Mallory Baker, Julie Kinsman, Lisa Mancl, Susan
  Norton, Randall Bly, and Shyamnath Gollakota.
\newblock Performing tympanometry using smartphones.
\newblock {\em Communications Medicine}, 2, 06 2022.

\bibitem{heitmann2021mass}
Nils Heitmann, Thomas Rosner, and Samarjit Chakraborty.
\newblock Mass-deployable smartphone-based objective hearing screening with
  otoacoustic emissions.
\newblock In {\em Proceedings of the 2021 International Conference on
  Multimodal Interaction}, pages 653--661, 2021.

\bibitem{heitmann2017sound4all}
Nils Heitmann, Philipp Kindt, Thomas Rosner, Kapil Sikka, Amit Chirom, Dinesh
  Kalyanasundaram, and Samarjit Chakraborty.
\newblock Sound4all: Towards affordable large-scale hearing screening.
\newblock In {\em 2017 12th International Conference on Design \& Technology of
  Integrated Systems In Nanoscale Era (DTIS)}, pages 1--6. IEEE, 2017.

\bibitem{pathmedical}
Path {M}edical - {A}udiology | {D}iagnostic | {S}creening | {T}racking.
\newblock 2021.
\newblock \url{https://www.pathme.de/}.

\bibitem{otoscreen1}
Otoglobal {H}ealth.
\newblock 2021.
\newblock \url{https://otoglobalhealth.wixsite.com/companysite}.

\bibitem{otoscreen2}
Novel pediatric hearing screening device.
\newblock 2018.
\newblock \url{http://jhu.technologypublisher.com/technology/40355}.

\bibitem{heitmann2020designing}
Nils Heitmann, Thomas Rosner, and Samarjit Chakraborty.
\newblock Designing a single speaker-based ultra low-cost otoacoustic emission
  hearing screening probe.
\newblock In {\em 2020 IEEE Global Humanitarian Technology Conference (GHTC)},
  pages 1--8. IEEE, 2020.

\bibitem{nurapatent}
Luke~John Campbell and Kyle~Damon Slater.
\newblock Personalization of auditory stimulus, Patent US10154333B2, Dec. 2018.

\bibitem{nura}
Nura.
\newblock 2021.
\newblock \url{https://www.nuraphone.com/}.

\bibitem{esense1}
Fahim Kawsar, Chulhong Min, Akhil Mathur, and Alessandro Montanari.
\newblock Earables for personal-scale behavior analytics.
\newblock {\em IEEE Pervasive Computing}, 17(3):83--89, 2018.

\bibitem{esense2}
Chulhong Min, Akhil Mathur, and Fahim Kawsar.
\newblock Exploring audio and kinetic sensing on earable devices.
\newblock In {\em Proceedings of the 4th ACM Workshop on Wearable Systems and
  Applications}, WearSys '18, page 5–10, New York, NY, USA, 2018. Association
  for Computing Machinery.

\bibitem{clearbuds}
Ishan Chatterjee, Maruchi Kim, Vivek Jayaram, Shyamnath Gollakota, Ira
  Kemelmacher, Shwetak Patel, and Steven~M. Seitz.
\newblock Clearbuds: Wireless binaural earbuds for learning-based speech
  enhancement.
\newblock In {\em Proceedings of the 20th Annual International Conference on
  Mobile Systems, Applications and Services}, MobiSys '22, page 384–396, New
  York, NY, USA, 2022. Association for Computing Machinery.

\bibitem{openearable}
Tobias Röddiger, Tobias King, Dylan~Ray Roodt, Christopher Clarke, and Michael
  Beigl.
\newblock Openearable: Open hardware earable sensing platform.
\newblock In {\em Proceedings of the 1st International Workshop on Earable
  Computing}, EarComp’22, page 29–34, New York, NY, USA, 2022. Association
  for Computing Machinery.

\bibitem{fda}
510(k) summary of safety and effectiveness: Titan with {DPQAE44O} and titan
  with {ABRIS440}.
\newblock 2022.
\newblock \url{https://www.accessdata.fda.gov/cdrh_docs/pdf10/K103760.pdf}.

\end{thebibliography}
